*Review*

# Cutting edge of high-entropy alloy superconductors from the perspective of materials research

**Jiro Kitagawa** [1,*], **Shusuke Hamamoto** [1], **and Naoki Ishizu** [1]

[1] Department of Electrical Engineering, Faculty of Engineering, Fukuoka Institute of Technology, 3-30-1 Wajiro-higashi, Higashi-ku, Fukuoka 811-0295, Japan

* Correspondence: j-kitagawa@fit.ac.jp

**Abstract:** High-entropy alloys (HEAs) are a new class of materials which are being energetically studied around the world. HEAs are characterized by a multi-component alloy in which five or more elements randomly occupy a crystallographic site. The conventional HEA concept has developed into simple crystal structures such as face-centered-cubic (fcc), body-centered-cubic (bcc) and hexagonal-closed packing (hcp) structures. The highly atomic-disordered state produces many superior mechanical or thermal properties. Superconductivity has been one of the topics of focus in the field of HEAs since the discovery of the bcc HEA superconductor in 2014. A characteristic of superconductivity is robustness against atomic disorder or extremely high pressure. The materials research on HEA superconductors has just begun, and there are open possibilities for unexpectedly finding new phenomena. The present review updates the research status of HEA superconductors. We survey bcc and hcp HEA superconductors and discuss the simple material design. The concept of HEA is extended to materials possessing multiple crystallographic sites; thus, we also introduce multi-site HEA superconductors with the CsCl-type, α-Mn-type, A15, NaCl-type, σ-phase and layered structures and discuss the materials research on multi-site HEA superconductors. Finally, we present the new perspectives of eutectic HEA superconductors and gum metal HEA superconductors.

## 1. Introduction

High-entropy alloys (HEAs) can be considered as a new class of materials especially due to their superior mechanical properties [1-4]. The conventional HEA concept was originally proposed for simple crystal structures such as face-centered-cubic (fcc), body-centered-cubic (bcc) and hexagonal-closed packing (hcp) structures, all of which possess only one crystallographic site. One of the definitions of HEA is that more than five elements with an atomic fraction of each element between 5% and 35% randomly occupy one crystallographic site [5]. The other definition follows the value of mixing entropy $\Delta S_{mix}$, which is expressed as follows:

$$\Delta S_{mix} = -R \sum_{i=1}^{n} c_i \ln c_i \tag{1}$$

where $n$ is the number of components, $c_i$ is the atomic fraction and $R$ is the gas constant. According to this equation, we classify low-entropy alloys as having an $R$ less than 0.69, medium-entropy alloys (MEAs) have an $R$ between 0.69 and 1.60 and HEAs have an $R$ of 1.60 or larger [6]. In this review, we basically follow the first definition for the single-site (fcc, bcc, and hcp) HEAs. The concept of HEA has now been adopted in many multi-site alloys beyond bcc, fcc and hcp structures. In such a case, a multi-component form is realized in the specific or respective crystallographic site in the material, hereafter called multi-site HEA. A quantitative definition of multi-site HEA does not exist; thus, we regard a reported compound as a multi-site HEA if the word "HEA" is used for the compound in the paper. Figure 1(a) draws the crystal structure of a single-site (bcc) HEA. In the bcc structure (space group: $Im\bar{3}m$, No. 229), only the 2$a$ site is present, which is randomly occupied by several atoms.

Figures 1(b) and 1(c) show the examples of multi-site HEAs with the A15 phase (space group: $Pm\bar{3}n$, No. 223), possessing sites 2*a* and 6*d*. In Fig. 1(b), the 6*d* site is randomly occupied by several different atoms, while the specific atom occupies the 2*a* site. On the other hand, both Wyckoff positions show an atomic disorder in Fig. 1(c).

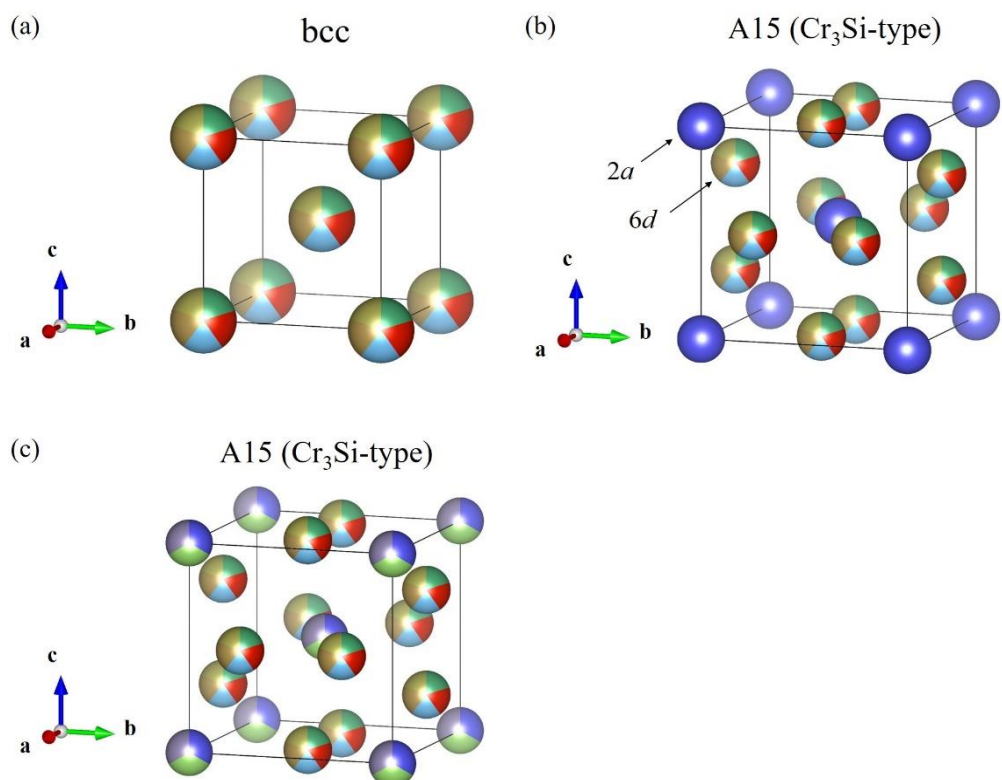

**Figure 1.** Crystal structures of (a) body-centered-cubic (bcc) and (b)(c) A15 compounds. The multi-colored balls mean that the site is randomly occupied by several atoms. The solid lines represent the unit cells.

HEAs show many superior properties such as a combination of high yield strength and ductility [7], high strength at high temperatures [8], strong resistance to corrosion and oxidation [9], outstanding thermal stability [10] and so on, which are primarily derived from the high atomic disorder. One of the attractive properties of HEAs is superconductivity. Since the discovery of a bcc HEA superconductor in 2014 [11], the number of reports of HEA superconductors is growing. Many HEA superconductors composed of transition metals can be regarded as an intermediate material between a transition metal crystalline superconductor and amorphous superconductor, from the perspective of the so-called Matthias rule. This rule is an empirical one, showing broad peak structures at the specified valence electron count per atom (VEC), when the superconducting critical temperature $T_c$ of transition metal crystalline superconductors is plotted as a function of VEC [12]. On the contrary, transition metal amorphous superconductors do not follow this rule and frequently show rather high $T_c$ values at the valley in the curve of the Matthias rule [13]. The VEC dependence of $T_c$ for HEA superconductors with a bcc structure seems to fall between the crystalline curve and the amorphous curve; thus, it is anticipated that HEA superconductors will be useful for the study of the relationship between crystalline and amorphous superconductors. Another important feature of HEA superconductivity is the robustness of $T_c$ against atomic disorder [14] or extremely high pressure [15]. New materials have provided many breakthroughs in materials science; thus, it is necessary to carry out materials research on HEA superconductors to pursue a new phenomenon. Although basic metallurgical research works are important for consistency, some hints regarding the material design are required, especially based on a comprehensive survey of the already reported materials.

In 2019, a review of HEA superconductors was published [16]. This review summarized the data of HEA superconductors with mainly bcc and hcp structures and presented a good deal of perspectives. After the publication, the research in this area has progressed, and several perspectives have been addressed. This review aims not only to update the research status of bcc and hcp HEA superconductors but also to present a survey of multi-site HEA superconductors. In each system, we aim to discuss the simple material design. Moreover, we present new perspectives in this research area. This review is organized as follows. In section 2, the bcc HEA superconductors, which are the most extensively studied of all types, are surveyed. Studies of hcp HEA superconductors are presented in section 3. In section 4, the simple material design of bcc and hcp HEA superconductors is discussed. Multi-site HEA superconductors are introduced in section 5. Several perspectives are given in section 6, and section 7 presents a summary of the review.

**2. bcc HEA superconductors**

*2.1. Ta-Nb-Hf-Zr-Ti*

The Ta-Nb-Hf-Zr-Ti system is the most explored HEA superconductor (see also numbers 1 to 13 and No. 21 in Table 1). The first HEA superconductor—$Ta_{34}Nb_{33}Hf_8Zr_{14}Ti_{11}$—was reported in 2014 [11]. It shows type II superconductivity at $T_c$ = 7.3 K, and the upper critical field $H_{c2}(0)$ is determined to be 82 kOe. The measured physical properties suggest a BCS-type phonon-mediated superconductor at the weak electron–phonon coupling limit.

After the discovery of $Ta_{34}Nb_{33}Hf_8Zr_{14}Ti_{11}$, extensive and systematic studies have been undertaken. The detailed VEC dependence of $T_c$ in $(TaNb)_{1-x}(HfZrTi)_x$ ($0.2 \leqq x \leqq 0.84$) was reported (No. 2 in Table 1) [14]. $T_c$ ranges from 4.5 K to 8.0 K, depending on $x$. A maximum $T_c$ is reached at approximately a VEC of 4.7, which mimics the feature of the Matthias rule for crystalline transition metal superconductors. The curve of the VEC dependence of $T_c$ for the HEAs falls between those of crystalline 4$d$ metal solid solutions and amorphous 4$d$ metals. Taking into account the fact that $Nb_{0.67}Ti_{0.33}$ is a $T_c$ = 9.2 K superconductor, the atomic disorder introduced in the HEA has a small influence on $T_c$. Furthermore, a research team has investigated the superconducting states of $(TaNb)_{0.67}(HfZrTi)_{0.33}$ substituted by isoelectronic mixtures of {Sc-Cr}, {Y-Mo} and {Sc-Mo} (No. 6 to No. 8 in Table 1) [17]. The $T_c$ of $(TaNb)_{0.67}(HfZrTi)_{0.33}$ was highly reduced by the replacement of Ta or Nb atoms; this means that $T_c$ strongly depends on the elemental makeup of the alloy. The team has also found that $(TaNb)_{0.67}(HfZrTi)_{0.33}Al_x$, which is the alloying of Al into $(TaNb)_{0.67}(HfZrTi)_{0.33}$, has a bcc structure up to $x$ = 0.3 (No. 9 in Table 1) [17]. The VEC dependence of $T_c$ also falls between the curves for crystalline and amorphous superconductors.

One of the high-impact results is the superconducting properties under high pressure [15]. The $(TaNb)_{0.67}(HfZrTi)_{0.33}$ HEA superconductor shows a robust zero-resistance against a high pressure of up to 190.6 GPa. This result is an important factor making superconducting HEAs a promising candidate for superconducting materials working under extreme conditions. Another important result to be noticed is the thermal annealing effect, as HEA superconductors are usually as-cast samples. The annealing effect has been reported by investigating the superconducting properties of Ta-Nb-Hf-Zr-Ti HEAs prepared under different thermal treatments [18]. Long-term annealing induces a short-range clustering of atoms, changing the microstructure of the HEA. However, the superconducting properties are rather insensitive to the microstructure difference.

In 2020, some pioneering studies have been reported. One presented the successful preparation of thin film HEA superconductors $(TaNb)_{1-x}(HfZrTi)_x$ using a magnetron sputtering method (No. 21 in Table 1) [19]. The films with $x$ < 0.65 showed normal metallic properties; on the other hand, those with $x$ > 0.65 exhibited weakly insulating behaviors. All films, except for $x$ = 0 and 1, entered into the superconducting states at low temperatures. The highest $T_c$ = 6.8 K was observed at $x$ = 0.43 with VEC = 4.57. The maximum $T_c$ was slightly lower than the value of the bulk sample ($T_c$ = 8 K, VEC = 4.7). This study paves the way for engineering research of HEA superconductors. The other pioneering work [20] was the evaluation of the critical current density $J_c$. The sample comprised the bcc HEA superconductor $Ta_{1/6}Nb_{2/6}Hf_{1/6}Zr_{1/6}Ti_{1/6}$ with $T_c$ = 7.9 K, which is characterized as a strong coupled s-wave superconductor with a dirty limit (No. 10 in Table 1). The zero-field $J_c$ is estimated to be 10,655 A cm$^{-2}$ at 2 K, which is inferior to that of the conventional binary alloy NbTi superconductor, exceeding 10$^5$ A cm$^{-2}$ at 4.2 K [21,22]. The vortex pinning force can be explained by the Dew-Hughes double exponential pinning model, which refers to the existence of two types of pinning mechanism. Another interesting result is the high Kadowaki–Woods ratio. Therefore, the HEA superconductor might be regarded as a strongly correlated system. The last paper to be highlighted is the report of superconductivity in uranium-containing HEA (No.11 in Table 1) [23]. In this report, Zr in (TaNb)$_{1-x}$(HfZrTi)$_x$ was replaced by a U atom. Generally, Zr and U possess similar chemical and electronic properties; for example, they can both adopt a tetravalent state with similar ionic sizes. Thus, the obtained bcc HEA superconductor (TaNb)$_{0.31}$(HfUTi)$_{0.69}$ was the first U-containing HEA superconductor. U is considered to be a magnetic atom; however, (TaNb)$_{0.31}$(HfUTi)$_{0.69}$ did not show a Curie–Weiss behavior, and 5$f$ electrons in U atoms would be itinerant.

## 2.2. The other systems

NbTaTiZr-based HEA superconductors supplemented with Fe, Ge, Si or V have been extensively investigated (No.1 4 to No. 18 in Table 1) [24]. NbTaTiZr enters into a superconducting state below $T_c$= 8.3 K. The addition of Ge and Ge+V enhance $T_c$ by up to 8.5 and 8.4 K, respectively. Microstructure analyses have revealed that a Nb-Ta rich phase plays an important role in increasing $T_c$. Besides this, $T_c$ seems to depend on the degree of lattice distortion. The equimolar HEA superconductor Nb$_{20}$Re$_{20}$Zr$_{20}$Hf$_{20}$Ti$_{20}$ is a type-II superconductor with a $T_c$ of 5.3 K and possesses an $H_{c2}(0)$ of 89 kOe (No.19 in Table 1) [25]. The temperature dependence of the specific heat can be well described by the single-gap BCS model. Hf$_{21}$Nb$_{25}$Ti$_{15}$V$_{15}$Zr$_{24}$ is a new bcc HEA superconductor with a $T_c$ of 5.3 K (No. 20 in Table 1) [26]. This study is inspired by the recent report [27] of equimolar bcc HfNbTiVZr, which shows a secondary phase after thermal annealing. Hf$_{21}$Nb$_{25}$Ti$_{15}$V$_{15}$Zr$_{24}$ was discovered by changing the atomic composition and is stable even after annealing.

**Table 1.** Superconducting properties, $\delta$, valence electron count (VEC) and e/a of bcc high-entropy alloy (HEA) superconductors. $T_c$ and $H_{c2}(0)$ data are from the references listed in the table.

| No. | HEA | $T_c$ (K) | $H_{c2}(0)$ (kOe) | $\delta$ | VEC | e/a | Ref. |
|---|---|---|---|---|---|---|---|
| 1 | Ta$_{34}$Nb$_{33}$Hf$_8$Zr$_{14}$Ti$_{11}$ | 7.3 | 82 | 4.56 | 4.67 | 1.44 | [11] |
| 2 | (TaNb)$_{1-x}$(HfZrTi)$_x$ (0.2≦x≦0.84) | | | | | | [14] |
| | x=0.3 | 8.0 | 67 | 4.34 | 4.7 | 1.45 | |
| | x=0.33 | 7.8 | 78 | 4.47 | 4.67 | 1.45 | |
| | x=0.4 | 7.6 | 84 | 4.71 | 4.6 | 1.45 | |
| | x=0.5 | 6.5 | 117 | 4.91 | 4.5 | 1.45 | |
| | x=0.84 | 4.5 | 90 | 4.64 | 4.16 | 1.46 | |
| 3 | (TaNbV)$_{0.67}$(HfZrTi)$_{0.33}$ | 4.3 | - | 6.39 | 4.67 | 1.33 | [14] |
| 4 | (NbV)$_{0.67}$(HfZrTi)$_{0.33}$ | 7.2 | - | 7.04 | 4.67 | 1.23 | [14] |
| 5 | (TaV)$_{0.67}$(HfZrTi)$_{0.33}$ | 4.0 | - | 7.04 | 4.67 | 1.31 | [14] |
| 6 | (TaNb)$_{0.67}$(HfZrTi)$_{0.33}$ substituted by {Sc-Cr} | | | | | | [17] |
| | (Sc$_{0.33}$Cr$_{0.67}$Nb)$_{0.67}$(HfZrTi)$_{0.33}$ | 5.6 | - | 9.14 | 4.67 | 1.28 | |
| | (TaSc$_{0.33}$Cr$_{0.67}$)$_{0.67}$(HfZrTi)$_{0.33}$ | 4.4 | - | 9.13 | 4.67 | 1.36 | |
| | (TaNb)$_{0.67}$(Sc$_{0.67}$Cr$_{0.33}$ZrTi)$_{0.33}$ | 7.5 | - | 5.68 | 4.67 | 1.39 | |
| | (TaNb)$_{0.67}$(HfSc$_{0.67}$Cr$_{0.33}$Ti)$_{0.33}$ | 7.4 | - | 5.39 | 4.67 | 1.42 | |
| | (TaNb)$_{0.67}$(HfZrSc$_{0.67}$Cr$_{0.33}$)$_{0.33}$ | 7.6 | - | 6.15 | 4.67 | 1.46 | |
| 7 | (TaNb)$_{0.67}$(HfZrTi)$_{0.33}$ substituted by {Y-Mo} | | | | | | [17] |
| | (Y$_{0.33}$Mo$_{0.67}$Nb)$_{0.67}$(HfZrTi)$_{0.33}$ | 4.7 | - | 8.98 | 4.67 | 1.44 | |
| | (TaY$_{0.33}$Mo$_{0.67}$)$_{0.67}$(HfZrTi)$_{0.33}$ | 3.5 | - | 8.96 | 4.67 | 1.53 | |

| | | | | | | |
|---|---|---|---|---|---|---|
| | (TaNb)$_{0.67}$(Y$_{0.67}$Mo$_{0.33}$ZrTi)$_{0.33}$ | 7.6 | - | 7.25 | 4.67 | 1.45 | |
| | (TaNb)$_{0.67}$(HfY$_{0.67}$Mo$_{0.33}$Ti)$_{0.33}$ | 6.7 | - | 7.06 | 4.67 | 1.48 | |
| | (TaNb)$_{0.67}$(HfZrY$_{0.67}$Mo$_{0.33}$)$_{0.33}$ | 7.5 | - | 7.47 | 4.67 | 1.51 | |
| 8 | (TaNb)$_{0.67}$(HfZrTi)$_{0.33}$ substituted by {Sc-Mo} | | | | | | [17] |
| | (Sc$_{0.33}$Mo$_{0.67}$Nb)$_{0.67}$(HfZrTi)$_{0.33}$ | 4.4 | - | 6.62 | 4.67 | 1.38 | |
| | (TaSc$_{0.33}$Mo$_{0.67}$)$_{0.67}$(HfZrTi)$_{0.33}$ | 2.9 | - | 6.61 | 4.67 | 1.47 | |
| | (TaNb)$_{0.67}$(Sc$_{0.67}$Mo$_{0.33}$ZrTi)$_{0.33}$ | 7.5 | - | 5.10 | 4.67 | 1.41 | |
| | (TaNb)$_{0.67}$(HfSc$_{0.67}$Mo$_{0.33}$Ti)$_{0.33}$ | 6.6 | - | 4.78 | 4.67 | 1.44 | |
| | (TaNb)$_{0.67}$(HfZrSc$_{0.67}$Mo$_{0.33}$)$_{0.33}$ | 7.5 | - | 5.58 | 4.67 | 1.47 | |
| 9 | (TaNb)$_{0.67}$(HfZrTi)$_{0.33}$Al$_x$ | | | | | | [17] |
| | $x = 0.15$ | 6.6 | - | 4.27 | 4.45 | 1.65 | |
| | $x = 0.3$ | 3.0 | - | 4.08 | 4.29 | 1.81 | |
| 10 | Ta$_{1/6}$Nb$_{2/6}$Hf$_{1/6}$Zr$_{1/6}$Ti$_{1/6}$ | 7.9 | 120 | 4.92 | 4.5 | 1.43 | [20] |
| 11 | (TaNb)$_{0.31}$(HfUTi)$_{0.69}$ | 3.2 | 64 | 4.17 | 4.31 | - | [23] |
| 12 | Nb$_{22.1}$Ta$_{26.3}$Ti$_{16.6}$Zr$_{15.5}$Hf$_{19.5}$ | 7.1 | 20 | 4.91 | 4.48 | 1.47 | [24] |
| 13 | Nb$_{21.5}$Ta$_{18.1}$Ti$_{15.9}$Zr$_{14.4}$Hf$_{16.6}$V$_{13.5}$ | 5.1 | 20 | 6.21 | 4.53 | 1.38 | [24] |
| 14 | NbTaTiZrFe | 6.9 | - | 8.04 | 5.2 | 1.31 | [24] |
| 15 | NbTaTiZrGe | 8.5 | - | 8.07 | 4.4 | 1.91 | [24] |
| 16 | NbTaTiZrGeV | 8.4 | - | 8.07 | 4.5 | 1.75 | [24] |
| 17 | NbTaTiZrSiV | 4.3 | - | 9.88 | 4.5 | 1.74 | [24] |
| 18 | NbTaTiZrSiGe | 7.4 | - | 10.7 | 4.33 | 2.26 | [24] |
| 19 | Nb$_{20}$Re$_{20}$Zr$_{20}$Hf$_{20}$Ti$_{20}$ | 5.3 | 89 | 5.86 | 4.8 | 1.42 | [25] |
| 20 | Hf$_{21}$Nb$_{25}$Ti$_{15}$V$_{15}$Zr$_{24}$ | 5.3 | - | 6.80 | 4.4 | 1.36 | [26] |
| 21 | (TaNb)$_{1-x}$(HfZrTi)$_x$ Thin film ($0.04 \leqq x \leqq 0.88$) | | | | | | [19] |
| | $x = 0.21$ | 6.2 | 58 | 3.82 | 4.79 | 1.45 | |
| | $x = 0.33$ | 6.4 | 71 | 4.47 | 4.67 | 1.45 | |
| | $x = 0.43$ | 6.8 | 88 | 4.79 | 4.57 | 1.45 | |
| | $x = 0.54$ | 6.1 | 99 | 4.96 | 4.46 | 1.45 | |
| | $x = 0.65$ | 5.6 | 110 | 4.97 | 4.35 | 1.46 | |
| | $x = 0.76$ | 4.6 | 104 | 4.84 | 4.24 | 1.46 | |
| | $x = 0.88$ | 2.8 | 62 | 4.51 | 4.12 | 1.46 | |

*2.3. Comparision of properties between bcc HEA and binary alloy superconductors*

HEAs often show a cocktail effect, which means an enhancement of properties beyond the simple mixture of those of constituent elements. Here, we briefly discuss the possibility of the cocktail effect in $T_c$, which is evaluated by $T_c^{obs}/T_c^{base}$, where $T_c^{obs}$ and $T_c^{base}$ are the experimental $T_c$ and $T_c$ obtained by averaging the $T_c$ values of the constituent elements weighted by each atomic percentage, respectively, following the method reported in [11]. $T_c^{base}$ plays a baseline role, and a larger $T_c^{obs}/T_c^{base}$ indicates a strong cocktail effect. Table 2 summarizes the $\Delta S_{mix}/R$, $T_c^{obs}$, $T_c^{base}$, and $T_c^{obs}/T_c^{base}$ of the bcc HEA and bcc binary alloy (Nb$_{66}$Ti$_{33}$ [14], Ta$_{50}$V$_{50}$ [28], Ti$_{20}$V$_{80}$ [28]) superconductors. $\Delta S_{mix}/R$ is calculated using equation (1). Considering that VEC largely affects the value of $T_c$, the $\Delta S_{mix}/R$ dependence of $T_c^{obs}/T_c^{base}$ is investigated by fixing VEC at 4.67, with which a great deal of data are obtained. VEC in Table 1 is calculated by

$$\text{VEC} = \sum_{i=1}^{n} c_i \text{VEC}_i \qquad (2)$$

where VEC$_i$ is the VEC of element $i$. Figure 2(a) shows the results compared to Nb$_{66}$Ti$_{33}$; it is difficult to clearly determine the existence of a cocktail effect in HEAs.

Next, we examined the $\Delta S_{mix}/R$ dependence of $\delta$, which is calculated by

$$\delta = 100 \times \sqrt{\sum_{i=1}^{n} c_i \left(1 - \frac{r_i}{\bar{r}}\right)^2} \quad (3)$$

where $r_i$ is the atomic radius of element $i$ and $\bar{r}$ is the composition-weighted average atomic radius (see also Table 1). The values of $r_i$ are taken from the works presented in [3] and [29]. The parameter $\delta$ is a measure of the degree of the atomic size difference among the constituent elements. The $\Delta S_{mix}/R$ dependence of $\delta$ is shown in Fig. 2(b), which shows a positive correlation. This means that the high-entropy state tends to stabilize the bcc structure even under the rather large mismatch of atomic radii among constituent elements. At the present stage, the high-entropy effect in superconductor seems to be better correlated to the structural properties than the electronic properties.

**Table 2.** $\Delta S_{mix}/R$, $T_c^{obs}$, $T_c^{base}$, and $T_c^{obs}/T_c^{base}$ of bcc HEA and binary alloy ($Nb_{66}Ti_{33}$, $Ta_{50}V_{50}$, $Ti_{20}V_{80}$) superconductors.

| No. | HEA/binary alloy | $\Delta S_{mix}/R$ | $T_c^{obs}$ (K) | $T_c^{base}$ (K) | $T_c^{obs}/T_c^{base}$ |
|---|---|---|---|---|---|
| 1 | $Ta_{34}Nb_{33}Hf_8Zr_{14}Ti_{11}$ | 1.453 | 7.3 | 4.72 | 1.55 |
| 2 | $(TaNb)_{1-x}(HfZrTi)_x$ $(0.2 \leqq x \leqq 0.84)$ | | | | |
| | $x=0.3$ | 1.426 | 8.0 | 4.93 | 1.62 |
| | $x=0.33$ | 1.461 | 7.8 | 4.73 | 1.65 |
| | $x=0.4$ | 1.528 | 7.6 | 4.28 | 1.77 |
| | $x=0.5$ | 1.589 | 6.5 | 3.63 | 1.78 |
| | $x=0.84$ | 1.473 | 4.5 | 1.41 | 3.21 |
| 3 | $(TaNbV)_{0.67}(HfZrTi)_{0.33}$ | 1.733 | 4.3 | 4.38 | 0.982 |
| 4 | $(NbV)_{0.67}(HfZrTi)_{0.33}$ | 1.461 | 7.2 | 5.00 | 1.44 |
| 5 | $(TaV)_{0.67}(HfZrTi)_{0.33}$ | 1.461 | 4.0 | 3.40 | 1.18 |
| 6 | $(TaNb)_{0.67}(HfZrTi)_{0.33}$ substituted by {Sc-Cr} | | | | |
| | $(Sc_{0.33}Cr_{0.67}Nb)_{0.67}(HfZrTi)_{0.33}$ | 1.674 | 5.6 | 3.23 | 1.73 |
| | $(TaSc_{0.33}Cr_{0.67})_{0.67}(HfZrTi)_{0.33}$ | 1.674 | 4.4 | 1.62 | 2.71 |
| | $(TaNb)_{0.67}(Sc_{0.67}Cr_{0.33}ZrTi)_{0.33}$ | 1.531 | 7.5 | 4.72 | 1.59 |
| | $(TaNb)_{0.67}(HfSc_{0.67}Cr_{0.33}Ti)_{0.33}$ | 1.531 | 7.4 | 4.67 | 1.58 |
| | $(TaNb)_{0.67}(HfZrSc_{0.67}Cr_{0.33})_{0.33}$ | 1.531 | 7.6 | 4.69 | 1.62 |
| 7 | $(TaNb)_{0.67}(HfZrTi)_{0.33}$ substituted by {Y-Mo} | | | | |
| | $(Y_{0.33}Mo_{0.67}Nb)_{0.67}(HfZrTi)_{0.33}$ | 1.674 | 4.7 | 3.44 | 1.37 |
| | $(TaY_{0.33}Mo_{0.67})_{0.67}(HfZrTi)_{0.33}$ | 1.674 | 3.5 | 1.83 | 1.91 |
| | $(TaNb)_{0.67}(Y_{0.67}Mo_{0.33}ZrTi)_{0.33}$ | 1.531 | 7.6 | 4.75 | 1.60 |
| | $(TaNb)_{0.67}(HfY_{0.67}Mo_{0.33}Ti)_{0.33}$ | 1.531 | 6.7 | 4.70 | 1.43 |
| | $(TaNb)_{0.67}(HfZrY_{0.67}Mo_{0.33})_{0.33}$ | 1.531 | 7.5 | 4.73 | 1.59 |
| 8 | $(TaNb)_{0.67}(HfZrTi)_{0.33}$ substituted by {Sc-Mo} | | | | |
| | $(Sc_{0.33}Mo_{0.67}Nb)_{0.67}(HfZrTi)_{0.33}$ | 1.674 | 4.4 | 3.44 | 1.28 |
| | $(TaSc_{0.33}Mo_{0.67})_{0.67}(HfZrTi)_{0.33}$ | 1.674 | 2.9 | 1.83 | 1.58 |
| | $(TaNb)_{0.67}(Sc_{0.67}Mo_{0.33}ZrTi)_{0.33}$ | 1.531 | 7.5 | 4.75 | 1.58 |
| | $(TaNb)_{0.67}(HfSc_{0.67}Mo_{0.33}Ti)_{0.33}$ | 1.531 | 6.6 | 4.71 | 1.40 |
| | $(TaNb)_{0.67}(HfZrSc_{0.67}Mo_{0.33})_{0.33}$ | 1.531 | 7.5 | 4.73 | 1.59 |
| 9 | $(TaNb)_{0.67}(HfZrTi)_{0.33}Al_x$ | | | | |
| | $x = 0.15$ | 1.658 | 6.6 | 4.27 | 1.55 |
| | $x = 0.3$ | 1.664 | 3.0 | 3.92 | 0.765 |
| 10 | $Ta_{1/6}Nb_{2/6}Hf_{1/6}Zr_{1/6}Ti_{1/6}$ | 1.561 | 7.9 | 4.03 | 1.96 |
| 11 | (TaNb)$_{0.31}$(HfUTi)$_{0.69}$ | 1.592 | 3.2 | 2.42 | 1.32 |
| 12 | Nb$_{22.1}$Ta$_{26.3}$Ti$_{16.6}$Zr$_{15.5}$Hf$_{19.5}$ | 1.591 | 7.1 | 3.41 | 2.08 |
| 13 | Nb$_{21.5}$Ta$_{18.1}$Ti$_{15.9}$Zr$_{14.4}$Hf$_{16.6}$V$_{13.5}$ | 1.780 | 5.1 | 3.69 | 1.38 |
| 14 | NbTaTiZrFe | 1.609 | 6.9 | 2.94 | 2.34 |
| 15 | NbTaTiZrGe | 1.609 | 8.5 | 2.94 | 2.89 |
| 16 | NbTaTiZrGeV | 1.792 | 8.4 | 3.34 | 2.51 |
| 17 | NbTaTiZrSiV | 1.792 | 4.3 | 3.34 | 1.29 |
| 18 | NbTaTiZrSiGe | 1.792 | 7.4 | 3.24 | 2.28 |
| 19 | Nb$_{20}$Re$_{20}$Zr$_{20}$Hf$_{20}$Ti$_{20}$ | 1.609 | 5.3 | 2.42 | 2.2 |
| 20 | Hf$_{21}$Nb$_{25}$Ti$_{15}$V$_{15}$Zr$_{24}$ | 1.586 | 5.3 | 3.30 | 1.61 |
| 21 | (TaNb)$_{1-x}$(HfZrTi)$_{x}$ Thin film (0.04 ≦ $x$ ≦ 0.88) | | | | |
| | $x = 0.21$ | 1.292 | 6.2 | 5.52 | 1.12 |
| | $x = 0.33$ | 1.461 | 6.4 | 4.73 | 1.35 |
| | $x = 0.43$ | 1.551 | 6.8 | 4.08 | 1.67 |
| | $x = 0.54$ | 1.602 | 6.1 | 3.37 | 1.81 |
| | $x = 0.65$ | 1.604 | 5.6 | 2.65 | 2.11 |
| | $x = 0.76$ | 1.552 | 4.6 | 1.93 | 2.70 |
| | $x = 0.88$ | 1.417 | 2.8 | 1.15 | 2.43 |
| 22 | Nb$_{66}$Ti$_{33}$ (VEC = 4.67) | 0.6365 | 9.2 | 6.34 | 1.45 |
| 23 | Ta$_{50}$V$_{50}$ (VEC = 5) | 0.6932 | 2.35 | 4.89 | 0.48 |
| 24 | Ti$_{20}$V$_{80}$ (VEC = 4.8) | 0.5004 | 7.5 | 4.32 | 1.74 |

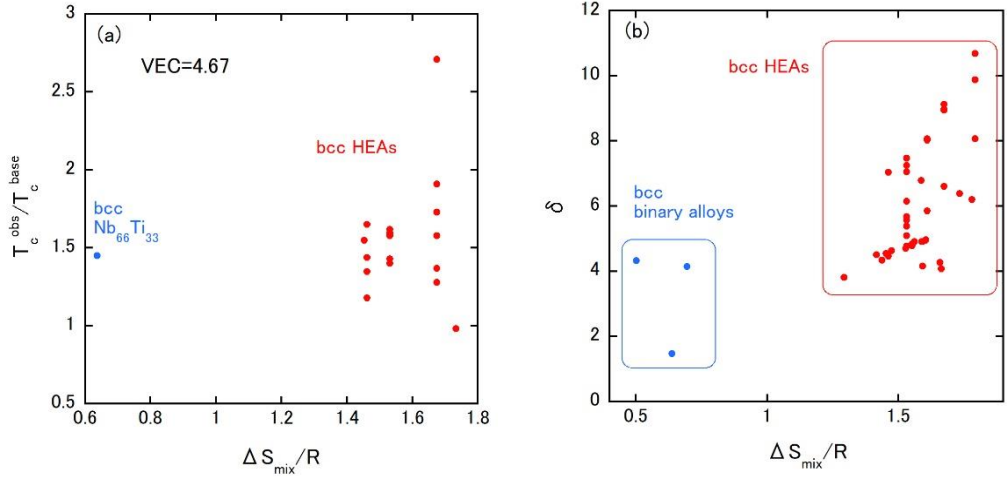

**Figure 2.** $\Delta S_{mix}/R$ dependence of (a) $T_c^{obs}/T_c^{base}$ and (b) $\delta$ for bcc HEA and binary alloy superconductors. The $\delta$-values of bcc binary alloys are 1.47 for Nb$_{66}$Ti$_{33}$, 4.15 for Ta$_{50}$V$_{50}$, and 4.33 for Ti$_{20}$V$_{80}$, respectively.

### 3. hcp and hcp-related HEA superconductors

To date, there has been only one report into hcp HEA superconductors (No. 1 in Table 3). We note here that, although the simple hcp (the Mg-type structure, $P6_3/mmc$, No.194) possesses only the 2c site, alloys No. 2 and No. 3 in Table 3 are multi-site HEAs with the same space group. In this review, alloys No. 2 and No. 3 are treated as hcp-related HEA superconductors.

Based on a $Mo_{33}Re_{33}Ru_{33}$ hcp alloy superconductor with a $T_c$ of 9.1 K, several hcp HEA or MEA superconductors are reported (see also No. 1 in Table 3) [30]. In this paper, $Mo_{33}Re_{33}Ru_{33}$ is alloyed by Ti and Rh elements, and $Mo_{.2375}Re_{.2375}Ru_{.2375}Rh_{.2375}Ti_{.05}$, $Mo_{.225}Re_{.225}Ru_{.225}Rh_{.225}Ti_{.1}$, $Mo_{.1}Re_{.1}Ru_{.55}Rh_{.1}Ti_{.15}$, $Mo_{.105}Re_{.105}Ru_{.527}Rh_{.105}Ti_{.158}$ and $Mo_{.118}Re_{.118}Ru_{.47}Rh_{.118}Ti_{.176}$ can be regarded as hcp HEA superconductors. The highest $T_c$ of 4.7 K was observed for $Mo_{.225}Re_{.225}Ru_{.225}Rh_{.225}Ti_{.1}$. Samples with a VEC lower than 7 tend to contain minor phases.

In $(Nb_{0.67-x}Re_x)(HfZrTi)_{0.33}$, as $x$ is increased from 0.1, a bcc to hexagonal structural transition occurs through the mixture of hcp and bcc [31]. $Re_{0.56}Nb_{0.11}Ti_{0.11}Zr_{0.11}Hf_{0.11}$ with an $x$ of 0.56 is a hexagonal HEA superconductor, which is characterized as a type-II superconductor with a $T_c$ of 4.4 K and $H_{c2}(0)$ of 36 kOe (No.2 in Table 3). In the HEA, there are three Wyckoff positions: $2a$, $6h$, and $4f$. Re and Nb atoms occupy the $2a$ and $6h$ sites, and Ti, Zr, and Hf atoms occupy the $4f$ site.

$Nb_{10+2x}Mo_{35-x}Ru_{35-x}Rh_{10}Pd_{10}$ ($0 \leqq x \leqq 5$) is also an hcp-related HEA superconductor (No. 3 in Table 3) [32]. Two Wyckoff positions $2a$ and $4f$ are randomly occupied by the constituent elements. With an increasing $x$, $T_c$ shows a maximum value of 6.2 K at a value of $x$ of 2.5, but $H_{c2}(0)$ maintains a monotonous increase and reaches 83 kOe at a value of $x$ of 5. From the specific heat results, it is suggested that these HEAs have a non BCS-like gap.

**Table 3.** Superconducting properties, $\delta$, VEC and e/a of hcp and hcp-related HEA superconductors. $T_c$ and $H_{c2}(0)$ data are from the references listed in the table.

| No. | HEA | $T_c$ (K) | $H_{c2}(0)$ (kOe) | $\delta$ | VEC | e/a | Ref. |
|---|---|---|---|---|---|---|---|
| 1 | Mo-Re-Ru-Rh-Ti | | | | | | [30] |
| | $Mo_{.2375}Re_{.2375}Ru_{.2375}Rh_{.2375}Ti_{.05}$ | 3.6 | - | 1.99 | 7.33 | 1.20 | |
| | $Mo_{.225}Re_{.225}Ru_{.225}Rh_{.225}Ti_{.1}$ | 4.7 | - | 2.54 | 7.15 | 1.20 | |
| | $Mo_{.1}Re_{.1}Ru_{.55}Rh_{.1}Ti_{.15}$ | 2.1 | - | 3.14 | 7.2 | 1.12 | |
| | $Mo_{.105}Re_{.105}Ru_{.527}Rh_{.105}Ti_{.158}$ | 2.2 | - | 3.19 | 7.16 | 1.13 | |
| | $Mo_{.118}Re_{.118}Ru_{.47}Rh_{.118}Ti_{.176}$ | 2.5 | - | 3.29 | 7.06 | 1.14 | |
| 2 | $Re_{0.56}Nb_{0.11}Ti_{0.11}Zr_{0.11}Hf_{0.11}$ | 4.4 | 36 | 6.00 | 5.79 | 1.41 | [31] |
| 3 | $Nb_{10+2x}Mo_{35-x}Ru_{35-x}Rh_{10}Pd_{10}$ | | | | | | [32] |
| | ($0 \leqq x \leqq 5$) | | | | | | |
| | $x = 0$ | 5.6 | 69 | 1.93 | 7.3 | 1.18 | |
| | $x = 2.5$ | 6.2 | 81 | 2.19 | 7.2 | 1.18 | |
| | $x = 5$ | 6.1 | 83 | 2.40 | 7.1 | 1.19 | |

## 4. Simple material design of bcc and hcp HEA superconductors

The comparison of $T_c$ vs VEC plots among three groups of crystalline $4d$ metal solid solutions, amorphous $4d$ metals and bcc HEAs is exhibited in Fig. 3(a). The HEAs are tentatively classified into Ta-Nb-Hf-Zr-Ti-based systems (Nos. 1–13 and No. 21 in Table 1), NbTaTiZr-based systems (Nos. 14–18 in Table 1) and the other group (No. 19 and No. 20 in Table 1). An HEA superconductor with a higher $T_c$ would be obtained at a VEC of 4.4–4.6. In the Ta-Nb-Hf-Zr-Ti-based systems, VEC dependences of $T_c$ are systematically investigated in detail (e.g., No. 2 and No. 21) and fall between the curves of crystalline and amorphous superconductors. It should be noted that, although the $T_c$ values of compounds No. 6 to No. 8 in Table 1 are distributed widely at the fixed VEC of 4.67, they are significantly affected by the elemental makeup, as mentioned below. Almost all HEA superconductors, including the alloys in the NbTaTiZr-based system and the other group, can be regarded as intermediate compounds between crystalline and amorphous compounds. Figure 3(b) shows the $\delta$ and VEC dependence of $T_c$ for the bcc HEA superconductors. The bcc structure is tolerable even for rather high $\delta$ values, and an interesting point is the robustness of $T_c$ against $\delta$ at a fixed VEC. This is the benefit of HEAs: they exhibit many superior properties due to their high mixing entropy, encouraging us to design a bcc HEA superconductor with a more flexible elemental makeup.

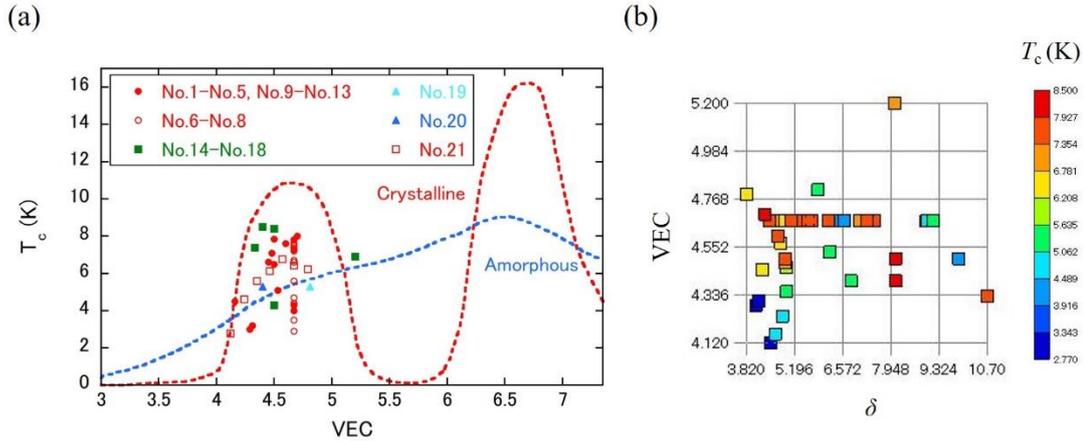

**Figure 3.** (a) VEC dependence of $T_c$ for bcc HEA superconductors. The same dependences are also shown for the crystalline 4$d$ metal solid solution and amorphous 4$d$ metal superconductors by the dotted lines. Data for the dotted lines are from [16]. (b) $\delta$ and VEC dependence of $T_c$ for bcc HEA superconductors.

When we design a bcc HEA superconductor, the elemental makeup is an important factor to be considered [17]. To elucidate the trend of the elemental makeup, the frequency with which elements are used in the bcc HEA superconductors reported to date (Nos. 1–20 in Table 1) is shown in Fig. 4 on the periodic table. The elemental makeup mainly consists of Hf, Zr, Ti, Ta, Nb, and V with a VEC of 4 or 5, due to the VEC requirement of the Matthias rule; that is, $T_c$ shows a broad peak at approximately a VEC of 4.5. A bcc HEA superconductor with a VEC larger than 4.8 is rarely found (see Fig. 3(a)), and to our knowledge, a bcc HEA superconductor with a VEC between 6 and 8 has not been reported. A VEC larger than 4.8 can be attained by using mainly Cr, Mo and/or W with the combination of 4$d$ or 5$d$ late-transition metals. In particular, a W-containing bcc HEA superconductor is still missing and would be a valuable topic of research. It should be noted here that, according to the well-known relation between VEC and the phase stability for fcc and bcc solid solutions in HEAs, a single bcc phase HEA would not be obtained at a VEC larger than 6.87 [1].

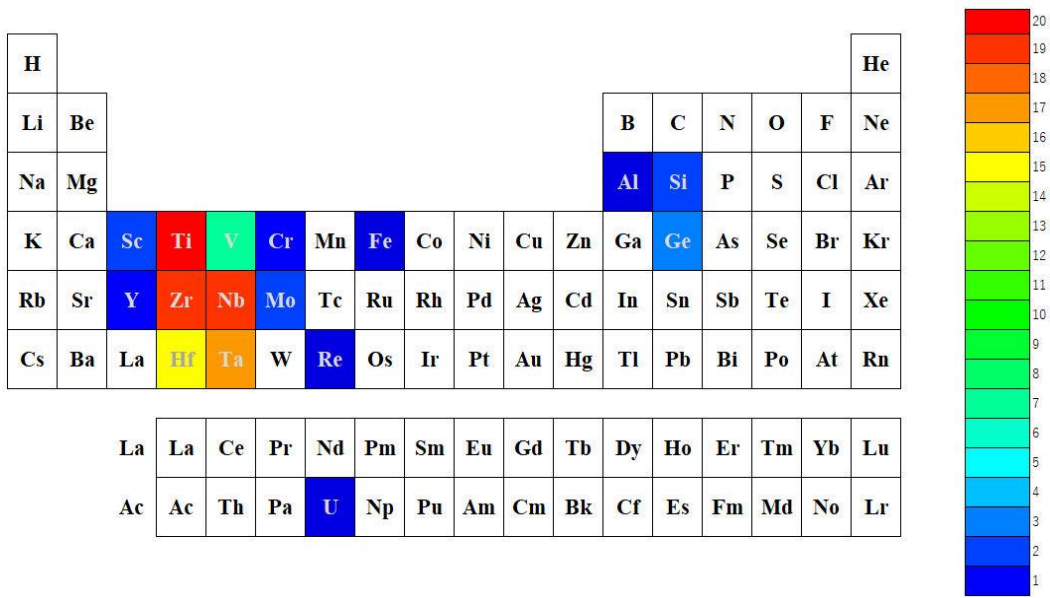

**Figure 4.** Frequency with which elements are used in bcc HEA superconductors reported to date (Nos. 1–20 in Table 1). The frequencies are 1 for Al, 2 for Si, 2 for Sc, 20 for Ti, 7 for V, 1 for Cr, 1 for Fe, 3 for Ge, 1 for Y, 19 for Zr, 19 for Nb, 2 for Mo, 15 for Hf, 17 for Ta, 1 for Re, and 1 for U.

Figure 5(a) displays the VEC dependence of $T_c$ for hcp and hcp-related HEA superconductors. The curves of crystalline and amorphous superconductors are also shown. Almost all HEAs show an enhancement of $T_c$ as the VEC is decreased to 7. In some alloy systems, samples with a VEC lower than 7 contain a secondary phase. Thus, an optimal VEC may be near 7.0. However, $T_c$ values tend to be lower than those of both crystalline and amorphous superconductors. $T_c$ is plotted as a function of $\delta$ and VEC in Fig. 5(b). As in the case of bcc HEA superconductors, $T_c$ is rather insensitive to $\delta$ for each system; however, the hcp HEAs seem to have little tolerance of $\delta$ compared to bcc HEAs. This may be an important factor for designing hcp HEAs. The relationship between VEC and the phase stability of fcc and bcc HEAs has been well established [1]: a single bcc phase for a VEC between 5.0 and 6.87 and a single fcc phase for a VEC larger than 8.0. On the other hand, the correlation of VEC and the hcp phase stability is not well understood, meaning that a wide range of VEC examination might be necessary for materials research. At present, almost all compounds concentrate on a narrow VEC range between 7.0 and 7.33, which is also contrasted with the wide distribution of VECs in the bcc HEA superconductors.

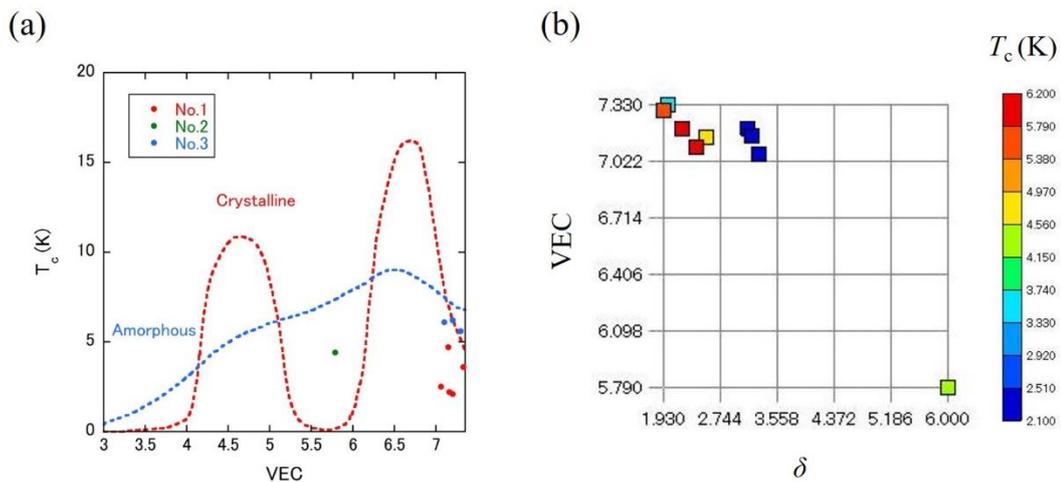

**Figure 5.** (a) VEC dependence of $T_c$ for hcp and hcp-related HEA superconductors. The same dependences are also shown for a crystalline 4$d$ metal solid solution and amorphous 4$d$ metal superconductors by the dotted lines. Data for the dotted lines are from [16]. (b) $\delta$ and VEC dependence of $T_c$ for hcp and hcp-related HEA superconductors.

The elemental makeup of hcp and hcp-related HEA superconductors is shown in Fig. 6, although the number of substances is small. In contrast to the bcc HEA superconductors, the constituent elements are equally distributed among Ti, Nb, Mo, Re, Ru and Rh with different VECs. The hcp elements of Re and Ru, in particular, might play an important role in the stabilization of the hcp structure. As can be seen from Fig. 4, a group 13 or 14 element can be added into the bcc HEA superconductor. In hcp HEA superconductors, the study of the addition of these kinds of elements is also needed.

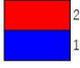

**Figure 6.** Frequency with which elements are used in hcp and hcp-related HEA superconductors reported to date (Nos. 1–3 in Table 3). The frequencies are 2 for Ti, 1 for Zr, 2 for Nb, 2 for Mo, 2 for Ru, 2 for Rh, 1 for Pd, 1 for Hf, and 2 for Re.

Here, we focus on both e/a and VEC to further elucidate the material design of bcc and hcp HEA superconductors. The value of e/a is the average number of itinerant electrons per atom and was originally proposed for the Hume–Rothery electron concentration rule. This concept is connected to the interaction of the Fermi surface with the Brillouin zone; thus, the e/a value is a good criterion for the stabilization of a crystal structure type [33]. On the other hand, VEC reflects the total density of states of the valence band integrated from the bottom up to a given energy. Therefore, VEC is a good scale of $T_c$, depending on the density of states at the Fermi level. The e/a-values of HEA superconductors are calculated as listed in Tables 1 and 3 using the updated e/a-value for each element [34]. Then, the phase selection diagram of bcc and hcp HEA superconductors based on VEC and e/a is constructed as shown in Fig. 7. While e/a values of bcc HEA superconductors are widely distributed, those of hcp superconductors might be in a narrow range, which means that the e/a of hcp HEA is useful for the judgment of phase stability. At the present stage, the phase selection diagram may work well, although VEC can solely distinguish between bcc and hcp HEA superconductors. The simultaneous consideration of e/a and VEC would assist in reliable material design.

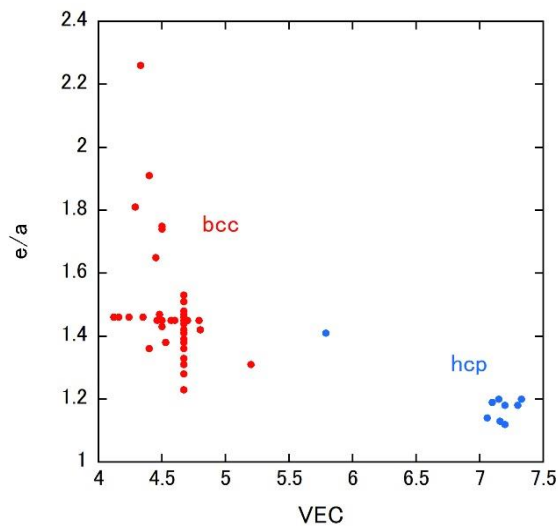

**Figure 7.** Phase selection diagram of bcc and hcp HEA superconductors. In hcp HEA superconductors, hcp-related superconductors are also included.

## 5. Multi-site HEA superconductors

### 5.1. CsCl

Pentanary (ScZrNb)$_{1-x}$(RhPd)$_x$ and hexanary (ScZrNbTa)$_{1-x}$(RhPd)$_x$ superconducting systems, both of which crystallize into the cubic CsCl-type structure, have been reported (No.1 in Table 4) [35]. The CsCl-type structure (space group: $Pm\bar{3}m$, No. 221) possesses the two inequivalent Wyckoff sites 1$a$ and 1$b$. The former and latter sites are randomly occupied by (Rh, Pd, and Nb) and (Nb, Sc, and Zr), respectively. The $T_c$ of (ScZrNb)$_{1-x}$(RhPd)$_x$ increases monotonously with decreasing VEC (see also Table 4) and does not follow the Matthias rule of the crystalline 4$d$ metal solid solutions. The highest $T_c$ of 9.5 K is observed in (ScZrNb)$_{0.65}$(RhPd)$_{0.35}$, which also presents the highest $H_{c2}(0)$ of 107 kOe.

### 5.2. α-Mn

(ZrNb)$_{1-x}$(MoReRu)$_x$, (HfTaWIr)$_{1-x}$Re$_x$ and (HfTaWPt)$_{1-x}$Re$_x$ form an α-Mn type structure in some $x$ ranges: $0.8 \leqq x \leqq 0.9$ for (ZrNb)$_{1-x}$(MoReRu)$_x$, $0.6 \leqq x \leqq 0.75$ for (HfTaWIr)$_{1-x}$Re$_x$ and $0.6 \leqq x \leqq 0.7$ for (HfTaWPt)$_{1-x}$Re$_x$, respectively (No. 2 in Table 4) [36]. The cubic α-Mn type structure with the space group of $I\bar{4}3m$ (No.217) is very unique, as it has a rather complex crystal structure with four Wyckoff positions (2$a$, 8$c$, 24$g$, 24$g$). In this structure, there are several binary alloy superconductors; therefore, the α-Mn type structure is a good platform for an HEA superconductor. The reported alloys are type-II superconductors, and the maximum $T_c$ values are found to be 5.3 K for (ZrNb)$_{1-x}$(MoReRu)$_x$ with an $x$ of 0.9, 5.6 K for (HfTaWIr)$_{1-x}$Re$_x$ with an $x$ of 0.75, and 5.7 K for (HfTaWPt)$_{1-x}$Re$_x$ with an $x$ of 0.7, respectively. The values of $T_c$ strongly depend on the cubic lattice parameter $a$ and VEC: there is a linearly increase of $T_c$ as $a$ decreases and VEC increases [36].

### 5.3. A15

A15 HEA (V$_{0.5}$Nb$_{0.5}$)$_{3-x}$Mo$_x$Al$_{0.5}$Ga$_{0.5}$ ($0.2 \leqq x \leqq 1.4$) shows superconductivity and temperature-induced polymorphism (No. 3 in Table 4) [37]. The crystal structure of cubic A15 is presented in Fig. 1(b) or Fig. 1(c). There are two Wyckoff sites: 2$a$ for Al and Ga and 6$d$ for V, Nb and Mo. The as-cast sample with an $x$ of 0.2 is a single bcc phase; however, upon annealing at 1600 °C, a polymorphic transformation to the A15 phase occurs. The other compositions are also A15 HEAs after annealing and exhibit superconductivity. It is interesting that, for an $x$ of 0.2, the A15 polymorph also shows a superconducting state at a $T_c$ of 10.2 K, while leaving the bcc polymorph normal state at 1.8 K. Both the $T_c$ and $H_{c2}(0)$ of the A15 HEAs decrease as the Mo content $x$ is increased (see also Table 4); however, the ratio of $H_{c2}(0)/T_c$ remains large, which suggests a disorder-induced enhancement of the upper critical field.

The difference in the synthesis route also affects the crystal structure produced. Fig. 8 shows our X-ray diffraction (XRD) patterns of Nb$_{2.8}$Mo$_{0.2}$Sn$_{0.8}$Al$_{0.2}$, denoted as #1 and #2, which are prepared by the arc melting plus thermal annealing at 800 °C and the solid-state reaction technique at 900 °C, respectively. The XRD pattern of #1 (#2) is well indexed by the bcc (A15) phase. Arc melting is usually performed at temperatures much higher than the solid-state reaction temperature; the higher reaction temperature leads to the dominance of the term including mixing entropy compared to the mixing enthalpy and can avoid the formation of an intermetallic compound such as the A15 phase, although Nb$_{2.8}$Mo$_{0.2}$Sn$_{0.8}$Al$_{0.2}$ with less atomic disorder might not have sufficiently high mixing entropy.

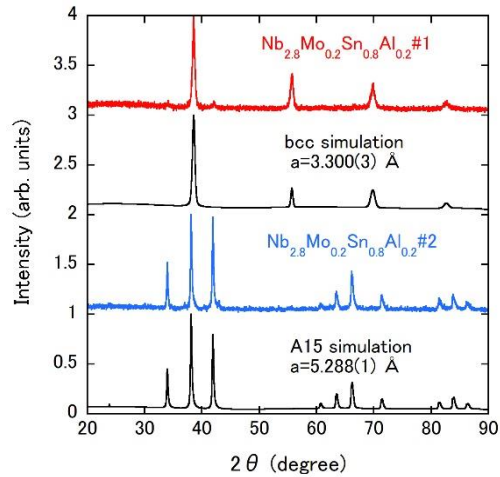

**Figure 8.** X-ray diffraction (XRD) patterns (Cu-K$\alpha$) of Nb$_{2.8}$Mo$_{0.2}$Sn$_{0.8}$Al$_{0.2}$ #1 and #2. The simulation patterns of bcc and A15 phases are also shown. Each pattern is shifted by an integer value for the clarity. The lattice parameters were determined with the help of the Rietveld refinement program [38,39].

*5.4. NaCl*

NaCl-type HEA AgInSnPbBiTe$_5$ shows superconductivity with a $T_c$ of 2.6 K (No. 4-1 in Table 4) [40]. In this structure (space group: $Fm\bar{3}m$, No. 225), the cation 4$b$ site is randomly occupied by Ag, In, Sn, Pb and Bi. A research group has also reported [41] the discovery of NaCl-type HEA superconductors of AgCdSnSbPbTe$_5$, AgInSnSbPbTe$_5$, AgCdInSnSbTe$_5$, AgCdSnPbBiTe$_5$, AgCdInPbBiTe$_5$ and AgCdInSnBiTe$_5$. The highest $T_c$ of 1.4 K was observed in AgInSnSbPbTe$_5$ (see also No. 4-2 in Table 4). A higher $T_c$ is reached for a telluride with a larger lattice constant. Recently, the same group has investigated the superconducting properties of a new HEA (Ag, In, Pb, Bi)Te$_{1-x}$Se$_x$ (No.4-3 in Table 4). The group has found that the superconducting state is robust against increased mixing entropy. To evaluate this, a new criterion of mixing entropy was introduced [42].

*5.5. σ-phase*

The tetragonal σ-phases (space group: $P4_2/mnm$, No. 136) composed of 4$d$ and 5$d$ elements often show superconductivity; for example, in Ta$_{1-x}$Re$_x$, Mo$_{1-x}$Ru$_x$, and W$_{1-x}$Ru$_x$ alloys. Two series of Ta$_5$(Mo$_{35-x}$W$_{5+x}$)Re$_{35}$Ru$_{20}$ ($0 \leqq x \leqq 30$) and (Ta$_{5+y}$Mo$_{35-y}$)W$_5$Re$_{35}$Ru$_{20}$ ($0 \leqq y \leqq 8$) HEAs are fully gapped σ-phase HEA superconductors (No. 5 in Table 4) [43]. The five Wyckoff positions (2$a$, 4$f$, 8$i$, 8$i$, 8$j$) of each HEA are randomly occupied by the constituent elements. The maximum $T_c$ is 6.3 K at $x = y =0$, which monotonously decreases as either $x$ or $y$ increases (see also Table 4). The $T_c$ vs VEC plot of the σ-phase HEAs falls on the curve of Matthias rule for the binary σ-phase superconductors.

*5.6. Layered superconductors*

REO$_{0.5}$F$_{0.5}$BiS$_2$ (RE: rare earth) is a BiS$_2$-based layered superconductor with an REO blocking layer. In [44,45], samples with different mixing entropies at the blocking layer have been prepared (No. 6-1 in Table 4). While the crystal structure parameters are almost independent of the mixing entropy, from PrO$_{0.5}$F$_{0.5}$BiS$_2$ with no mixing entropy to (La$_{0.2}$Ce$_{0.2}$Pr$_{0.2}$Nd$_{0.2}$Sm$_{0.2}$)O$_{0.5}$F$_{0.5}$BiS$_2$ with the highest mixing entropy, the bulk nature of superconductivity is improved. This suggests that the mixing entropy at the blocking layer severely affects the superconducting state. The other example is the RE123 HEA superconductors Y$_{0.28}$Nd$_{0.16}$Sm$_{0.18}$Eu$_{0.18}$Gd$_{0.20}$Ba$_2$Cu$_3$O$_{7-\delta}$ and Y$_{0.18}$La$_{0.24}$Nd$_{0.14}$Sm$_{0.14}$Eu$_{0.15}$Gd$_{0.15}$Ba$_2$Cu$_3$O$_{7-\delta}$, with $T_c$ values exceeding 90 K (No.6-2 in Table 4) [46]. The mixing entropy at the RE site, which is sandwiched between superconducting layers, does not change $T_c$ and $J_c$ as much. RE123 compounds with an orthorhombic structure show superconductivity, and $T_c$ is well correlated with the orthorhombicity. The $J_c$ of the HEA sample is larger than that of the conventional sample. Therefore, RE123 HEA would be useful in applications as a superconducting wire.

**Table 4.** Superconducting properties, $\delta$, VEC and e/a of multi-site HEA superconductors. $T_c$ and $H_{c2}(0)$ data are from the references listed in the table.

| No. | HEA | $T_c$ (K) | $H_{c2}(0)$ (kOe) | $\delta$ | VEC | e/a | Ref. |
|---|---|---|---|---|---|---|---|
| 1 | CsCl-type | | | | | | [35] |
| | $(ScZrNbTa)_{0.67}(RhPd)_{0.33}$ | 4.0 | 21 | 7.58 | 5.98 | 1.28 | |
| | $(ScZrNbTa)_{0.685}(RhPd)_{0.315}$ | 6.3 | 88 | 7.57 | 5.90 | 1.29 | |
| | $(ScZrNb)_{0.60}(RhPd)_{0.40}$ | 4.7 | 21 | 8.15 | 6.2 | 1.22 | |
| | $(ScZrNb)_{0.62}(RhPd)_{0.38}$ | 7.9 | 89 | 8.13 | 6.09 | 1.23 | |
| | $(ScZrNb)_{0.63}(RhPd)_{0.37}$ | 8.7 | 96 | 8.12 | 6.04 | 1.23 | |
| | $(ScZrNb)_{0.65}(RhPd)_{0.35}$ | 9.5 | 107 | 8.08 | 5.93 | 1.24 | |
| 2 | $\alpha$-Mn-type | | | | | | |
| 2-1 | $(ZrNb)_{1-x}(MoReRu)_x$ | | | | | | [36] |
| | $(0.8 \leqq x \leqq 0.9)$ | | | | | | |
| | $x = 0.8$ | 4.2 | - | 5.40 | 6.5 | 1.30 | |
| | $x = 0.9$ | 5.3 | 79 | 4.10 | 6.75 | 1.29 | |
| 2-2 | $(HfTaWIr)_{1-x}Re_x$ | | | | | | [36] |
| | $(0.6 \leqq x \leqq 0.75)$ | | | | | | |
| | $x = 0.6$ | 4.0 | 47 | 4.47 | 6.6 | 1.48 | |
| | $x = 0.7$ | 4.5 | - | 3.95 | 6.7 | 1.46 | |
| | $x = 0.75$ | 5.6 | - | 3.65 | 6.75 | 1.45 | |
| 2-3 | $(HfTaWPt)_{1-x}Re_x$ | | | | | | [36] |
| | $(0.6 \leqq x \leqq 0.7)$ | | | | | | |
| | $x = 0.6$ | 4.4 | 59 | 4.36 | 6.7 | 1.48 | |
| | $x = 0.7$ | 5.7 | - | 3.88 | 6.78 | 1.46 | |
| 3 | A15 | | | | | | [37] |
| | $(V_{0.5}Nb_{0.5})_{3-x}Mo_xAl_{0.5}Ga_{0.5}$ | | | | | | |
| | $(0.2 \leqq x \leqq 1.4)$ | | | | | | |
| | $x = 0.2$ | 10.2 | 201 | 3.72 | 4.55 | 1.60 | |
| | $x = 0.4$ | 9.2 | 177 | 3.61 | 4.6 | 1.61 | |
| | $x = 0.6$ | 8.9 | 170 | 3.51 | 4.65 | 1.63 | |
| | $x = 0.8$ | 7.9 | 142 | 3.40 | 4.7 | 1.64 | |
| | $x = 1.0$ | 6.1 | 99 | 3.28 | 4.75 | 1.65 | |
| | $x = 1.2$ | 4.8 | 76 | 3.16 | 4.8 | 1.67 | |
| | $x = 1.4$ | 3.2 | 48 | 3.04 | 4.85 | 1.68 | |
| 4 | NaCl | | | | | | |
| 4-1 | $AgInSnPbBiTe_5$ | 2.6 | 19 | 7.01 | 5.7 | - | [40] |
| 4-2 | $Ag_{0.20}Cd_{0.20}Sn_{0.20}Sb_{0.15}Pb_{0.20}Te_{1.05}$ | 1.2 | - | 6.54 | 6.63 | - | [41] |
| | $Ag_{0.24}In_{0.22}Sn_{0.18}Sb_{0.14}Pb_{0.19}Te_{1.03}$ | 1.4 | - | 7.05 | 5.83 | - | |
| | $Ag_{0.22}Cd_{0.22}In_{0.23}Sn_{0.17}Sb_{0.14}Te_{1.02}$ | 0.7 | - | 5.30 | 6.63 | - | |
| | $Ag_{0.19}Cd_{0.19}Sn_{0.20}Pb_{0.18}Bi_{0.21}Te_{1.03}$ | 1.0 | - | 6.43 | 6.56 | - | |
| | $Ag_{0.21}Cd_{0.19}In_{0.25}Pb_{0.16}Bi_{0.18}Te_{1.00}$ | 1.0 | - | 6.71 | 6.44 | - | |
| | $Ag_{0.21}Cd_{0.21}In_{0.24}Sn_{0.19}Bi_{0.19}Te_{0.97}$ | 1.0 | - | 5.49 | 6.54 | - | |
| 4-3 | $Ag_{0.24}In_{0.22}Pb_{0.27}Bi_{0.26}Te_{1.02}$ | 2.7 | 18 | 7.40 | 5.9 | - | [42] |
| | $Ag_{0.29}In_{0.26}Pb_{0.22}Bi_{0.24}Te_{0.78}Se_{0.20}$ | 2.5 | 19 | 7.58 | 5.97 | - | |
| | $Ag_{0.34}In_{0.15}Pb_{0.24}Bi_{0.29}Te_{0.65}Se_{0.34}$ | 2.0 | | 7.68 | 6.27 | - | |
| 5 | $\sigma$-phase | | | | | | |
| 5-1 | $Ta_5(Mo_{35-x}W_{5+x})Re_{35}Ru_{20}$ | | | | | | [43] |
| | $(0 \leq x \leq 30)$ | | | | | | |
| | $x = 0$ | 6.3 | - | 1.45 | 6.7 | 1.33 | |
| | $x = 5$ | 6.2 | - | 1.44 | 6.7 | 1.34 | |
| | $x = 10$ | 6.1 | - | 1.44 | 6.7 | 1.34 | |
| | $x = 15$ | 5.7 | - | 1.44 | 6.7 | 1.34 | |
| | $x = 20$ | 5.5 | - | 1.44 | 6.7 | 1.34 | |
| | $x = 25$ | 5.5 | - | 1.44 | 6.7 | 1.34 | |
| | $x = 30$ | 4.8 | - | 1.44 | 6.7 | 1.35 | |
| 5-2 | $(Ta_{5+y}Mo_{35-y})W_5Re_{35}Ru_{20}$ | | | | | | [43] |
| | $(0 \leq y \leq 8)$ | | | | | | |
| | $y = 2$ | 6.1 | - | 1.59 | 6.68 | 1.34 | |
| | $y = 4$ | 5.7 | - | 1.71 | 6.66 | 1.34 | |
| | $y = 6$ | 5.3 | - | 1.82 | 6.64 | 1.35 | |
| | $y = 8$ | 5.3 | - | 1.92 | 6.62 | 1.35 | |
| 6 | Layered superconductor | | | | | | |
| 6-1 | $(La_{0.2}Ce_{0.2}Pr_{0.2}Nd_{0.2}Sm_{0.2})O_{0.5}F_{0.5}BiS_2$ | 4.3 | - | - | 5.3 | - | [44,45] |
| | $(La_{0.3}Ce_{0.3}Pr_{0.2}Nd_{0.1}Sm_{0.1})O_{0.5}F_{0.5}BiS_2$ | 3.4 | - | - | 5.3 | - | |
| | $(La_{0.1}Ce_{0.1}Pr_{0.3}Nd_{0.3}Sm_{0.2})O_{0.5}F_{0.5}BiS_2$ | 4.7 | - | - | 5.3 | - | |
| | $(La_{0.1}Ce_{0.1}Pr_{0.2}Nd_{0.3}Sm_{0.3})O_{0.5}F_{0.5}BiS_2$ | 4.9 | - | - | 5.3 | - | |
| 6-2 | $(Y_{0.28}Nd_{0.16}Sm_{0.18}Eu_{0.18}Gd_{0.20})Ba_2Cu_3O_{7-\delta}$ | 93 | - | 46.7 | 6.31 | - | [46] |
| | $(Y_{0.18}La_{0.24}Nd_{0.14}Sm_{0.14}Eu_{0.15}Gd_{0.15})Ba_2Cu_3O_{7-\delta}$ | 93 | - | 46.8 | 6.31 | - | |

*5.7. Materials research based on δ, VEC and e/a*

Although the VEC values of some compounds might be meaningless, in Fig. 9(a), we have constructed the same plot as that of the bcc or hcp HEAs, except for compound No. 6 in Table 4, which does not contain the information concerning the VEC dependence of $T_c$. The A15 HEAs follow the Matthias rule. The CsCl-type and the NaCl-type HEAs show a deviation from the Matthias rule of crystalline 4$d$ metal solid solutions; however, in each system, a correlation between $T_c$ and VEC is observed, as $T_c$ tends to be enhanced as VEC decreases. While $T_c$ also seems to depend on VEC in the α-Mn-type or the σ-phase HEAs, the $T_c$ values are smaller than those of crystalline 4$d$ metal solid solutions and amorphous 4$d$ metals. The existence of a dependence of VEC on $T_c$ suggests the important role of the density of states for determining $T_c$. However, a universal dependence cannot be confirmed, because the shape of the density of states is uniquely governed by the crystal structure and/or the picture of a rigid band model may be broken in some cases.

Figure 9(b) shows the δ and VEC dependence of $T_c$ for HEA superconductors No. 1 to No. 5 in Table 4. Due to the lack of atomic radius values for the fluorine atom and the extraordinary large δ values of the RE123 system, compound No. 6 in Table 4 is not plotted. Interestingly, the $T_c$ of each system is insensitive to the magnitude of δ, which means that superconductivity lends robustness against δ. The δ-value indicates the degree of difference of the atomic radius between the constituent elements. A larger δ-value (e.g., >10) generally indicates the stabilization of an intermetallic compound. Therefore, it is expected that a multi-site HEAs tend to show a larger δ; however, except for the RE123 layered compounds, the δ-values are unexpectedly small. In the layered structures, a different function is assigned for each layer: for example, electrical conduction or charge supply. Even in this case, the δ-value in each layer would not be as large; e.g. an δ of 5.56 in the RE site in

($Y_{0.28}Nd_{0.16}Sm_{0.18}Eu_{0.18}Gd_{0.20}$)$Ba_2Cu_3O_{7-\delta}$ (No. 6-2 in Table 4). In other words, the multi-site HEAs No. 1 to No .5 in Table 4 have no different function for the respective sites, and all sites contribute to the superconductivity. We note that the ionic or covalent radii would be adequate for describing the differences of atomic species in the layered compounds, and the comparison of $\delta$-values between simple alloy structures and the layered structures might be meaningless. Further study is needed to address this issue. There is no information about a multi-site HEA with a larger $\delta$ for simple alloy structures, and it would be interesting to carry out materials research in this area.

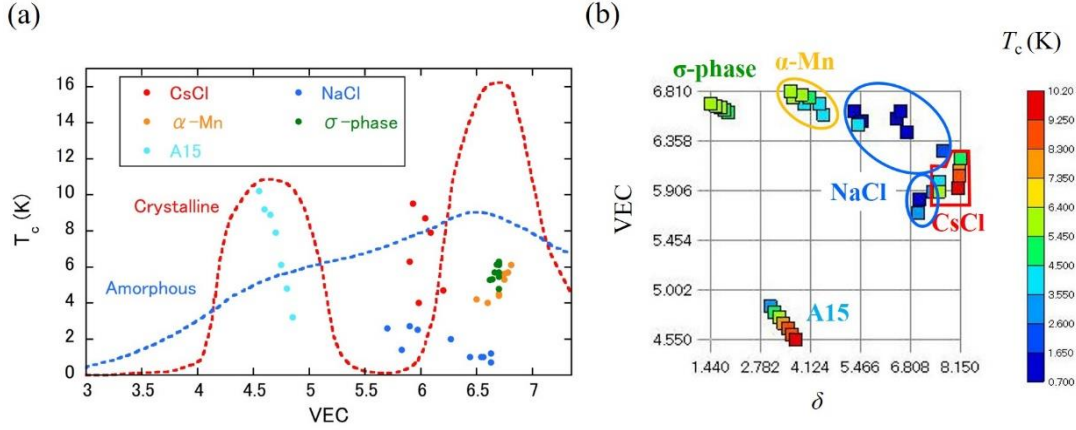

**Figure 9.** (a) VEC dependence of $T_c$ for multi-site HEA superconductors. The same dependences are also shown for crystalline and amorphous superconductors by the dotted lines. Data for the dotted lines are from [16]. (b) $\delta$ and VEC dependence of $T_c$ for multi-site HEA superconductors.

Finally, we comment on e/a, which is listed in Table 4 for each compound. Since the e/a value is not evaluated for oxygen, sulfur, selenium, tellurium or fluorine atoms, e/a values for compounds No. 4 and No. 6 are not given. Each crystal structure type has a narrow specific e/a range, making it possible to use for the phase selection. For instance, (HfTaWIr)$_{1-x}$Re$_x$ of the $\alpha$-Mn-type and (Ta$_{5+y}$Mo$_{35-y}$)W$_5$Re$_{35}$Ru$_{20}$ with the $\sigma$-phase have similar VEC values, but they can be separated by their e/a-values. It would be useful to also employ e/a for the design of multi-site HEAs.

## 6. Perspectives

The pioneering review [16] of HEA superconductors proposed a thin film HEA and the need for the evaluation of $J_c$, which have both recently been realized [19,20]. These kinds of research will be constantly carried out. The recent wave of machine learning has accelerated the research into the area of HEAs [47,48]. Although the number of HEA superconductors is rather small, a machine learning method will be promising for screening new HEA superconductors. In this review, we have selected five topics, including two new perspectives (6.4 and 6.5), as mentioned below.

*6.1. fcc HEA superconductor*

A review of HEA superconductors [16] proposed a possible fcc HEA superconductor. Although fcc-related NaCl-type HEA superconductors have been discovered [40-42], a simple fcc HEA superconductor has not yet been found. An fcc HEA superconductor would be useful to deepen our understanding of HEAs and/or of the relationship between crystalline and amorphous superconductors. We have carried out materials research on fcc HEA superconductors and employed the rather high $T_c$ element Nb in the chemical composition because the previously reported HEA superconductors contain superconducting elements [49]. Figure 10 shows the XRD patterns of some Nb-containing samples with VEC values larger than 8.0, which possess dominant fcc phases. These samples do not show superconducting signals down to 3 K. A single fcc phase of HEA would require a VEC larger than 8.0 [1]. According to the Matthias rule of crystalline 4$d$ metal solid solutions, the $T_c$ of a superconductor with a VEC larger than 8 is extremely low; thus, the materials research on a high $T_c$ fcc HEA superconductor may be unrealistic, but nevertheless represents a challenging theme.

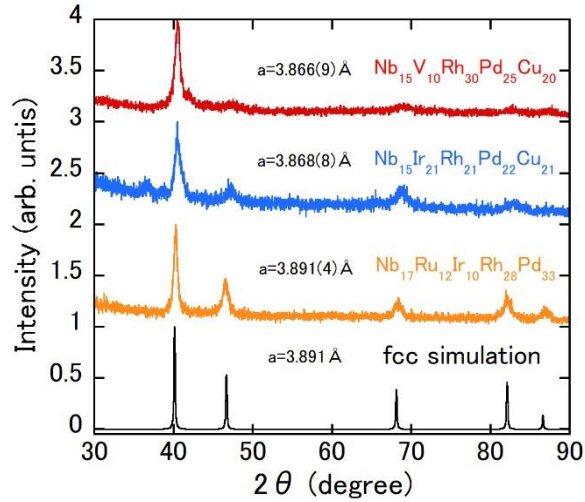

**Figure 10.** XRD patterns (Cu-K$\alpha$) of Nb$_{15}$V$_{10}$Rh$_{30}$Pd$_{25}$Cu$_{20}$, Nb$_{15}$Ir$_{21}$Rh$_{21}$Pd$_{22}$Cu$_{21}$ and Nb$_{17}$Ru$_{12}$Ir$_{10}$Rh$_{28}$Pd$_{33}$. The simulation pattern of the fcc phase with $a$ = 3.891 Å is also shown. Each pattern is shifted by an integer value for clarity. The lattice parameters were determined with the help of the Rietveld refinement program [38,39].

*6.2. Multi-site HEA superconductor*

Interest in multi-site HEA superconductors will grow due to the high degree-of-freedom of the HEA design. We have carried out materials research on the Mn$_5$Si$_3$-type HEA. Several superconductors with a hexagonal Mn$_5$Si$_3$-type—or its ordered derivative Ti$_5$Ga$_4$-type—structure have been reported [50-56]. Furthermore, each crystal structure type possesses rich intermetallic compounds [57,58]. The selected XRD patterns are given in Figure 11, showing the existence of HEA alloys for this structure type. The chemical compositions of Mn$_5$Si$_3$-type HEA #1 and #2 are (Sc$_{0.12}$Zr$_{0.18}$Ti$_{0.18}$Nb$_{0.24}$V$_{0.28}$)$_5$(Ga$_{0.13}$Ge$_{0.43}$Si$_{0.43}$)$_3$ and (Sc$_{0.25}$Zr$_{0.25}$Ti$_{0.25}$Nb$_{0.25}$)$_5$(Ge$_{0.6}$Si$_{0.4}$)$_3$, respectively. The ac magnetization measurement of Mn$_5$Si$_3$-type HEA #2 does not show a superconducting signal down to 3 K. Although a diamagnetic signal was observed in the Mn$_5$Si$_3$-type HEA #1 at approximately 3.8 K, it is due to a minor phase (Nb-V-based alloy) contained in the sample.

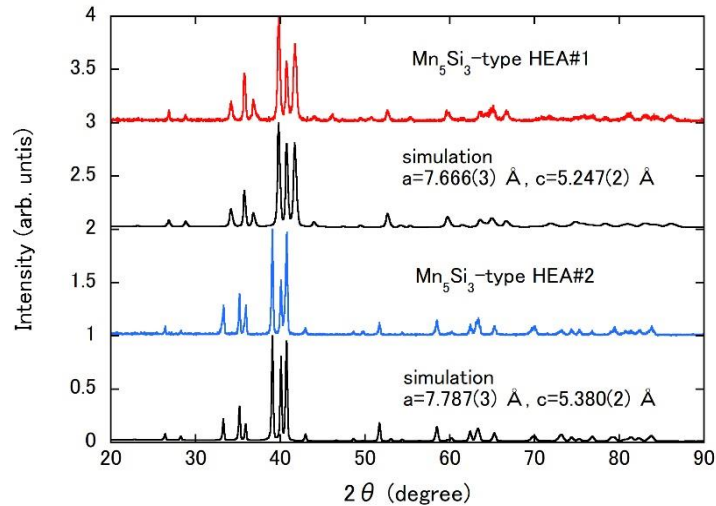

**Figure 11.** XRD patterns (Cu-K$\alpha$) of Mn$_5$Si$_3$-type HEAs. The simulation patterns are also shown. Each pattern is shifted by an integer value for clarity. The lattice parameters were determined with the help of the Rietveld refinement program [38,39].

*6.3. HEA superconductor containing magnetic element*

A previous review [16] has also proposed a HEA superconductor containing a magnetic element. Actually, $(TaNb)_{0.67}(HfZrTi)_{0.33}$ substituted by {Sc-Cr} or NbTaTiZrFe may be regarded as such a superconductor (see also Table 1), and detailed studies into this compound are desired. There seem to be two directions of research: one is the magnetic impurity effect in a HEA superconductor. The effect of magnetic impurity on superconductivity has a long history with superconducting elements, alloys and intermetallic compounds [59,60], and magnetic impurity usually destroys the Cooper pairs. Abrikosov–Gor'kov (A–G) pair-breaking theory describes the magnetic impurity concentration dependence of $T_c$ well. It would be interesting to determine whether HEA superconductors with magnetic impurities also follow the A–G theory or not. The other research direction is the search for HEA magnetic superconductors in which the magnetic element is cooperative with the superconductivity. In this case, we may anticipate an exotic superconducting state originating from the large mixing entropy.

*6.4. Eutectic HEA superconductor*

There is growing interest in eutectic HEAs due to the rich functions arising from the microstructures [61]. In superconductors, the microstructure containing the eutectic phase often contributes to the enhancement of $T_c$. The eutectic $Sr_2RuO_4$ sample shows a lamellar pattern of Ru metal, and $T_c$ increases from 1.5 K to 3 K [62]. Ir and a small amount of $YIr_2$ also form a lamellar pattern, which possesses a small lattice mismatch, leading to strain-induced lattice softening [63]. This softening causes the $T_c$ to increase from 0.1 K to 2.7 K. $Zr_5Pt_3O_x$ shows an interesting dependence of $x$ on $T_c$ [55]. Although the $T_c$ of 6.4 K in $Zr_5Pt_3$ is monotonously reduced to 3.2 K with $x$ increasing to 0.6, as $x$ is further increased from 0.6 to 2.5, $T_c$ increases to 4.8 K at an $x$ of 2.5. The metallographic investigation has revealed a change of microstructure at an $x$ of 1.0; the eutectic phase composed of $Zr_5Pt_3O_{0.5-0.6}$ and ZrPt appears and the area increases with an increase of $x$. The change of microstructure would be responsible for the enhancement of superconductivity. Moreover, a eutectic alloy often contributes to the improvement of $J_c$ [64]. The enhancement of $T_c$ and/or the improvement of $J_c$ are also expected for eutectic HEA superconductors. Recently, a eutectic phase has been reported [65] in Si-added bcc NbMoTiV. Since bcc HEA superconductors are extensively studied, the report of the eutectic phase in the bcc structure is highly encouraging. The practical superconducting wire employs a multifilamentary structure [66]. The microstructure of the eutectic superconductor may be regarded as a built-in multifilamentary structure; therefore, a eutectic HEA superconductor would have potential as a high-performance superconducting-wire.

*6.5. Gum metal HEA superconductor*

A specific class of β-titanium alloys after cold rolling shows unusual mechanical behaviors such as superelasticity and a low modulus; these are called gum metals, which are in practical use in wire frames for glasses, medical equipment and so on [67]. This material is attracting research attention worldwide in the field of titanium alloys. Gum metals are characterized by three specific values: a VEC of 4.24, a bond-order of 2.87 based on the DV-X$\alpha$ method and a $d$-electron orbital energy level of 2.45 eV. The VEC value is appropriate for the appearance of the superconductivity of $d$-electron alloy superconductors. Besides this, the compositions of gum metals have some implied similarities with HEAs. Figure 12 shows the preliminary results of the temperature dependences of the ac magnetic susceptibility $\chi_{ac}$(T) and electrical resistivity $\rho$(T) of as-cast $Al_5Nb_{24}Ti_{40}V_5Zr_{26}$, which was recently reported to be a gum metal-like HEA alloy [68]. The diamagnetic signal of $\chi_{ac}$ and the zero resistivity below approximately 5 K indicate superconductivity. Gum metals are highly advantageous for making wires; thus, if good superconducting properties are preserved even after cold rolling, this kind of material would be a good candidate for next-generation superconducting wire.

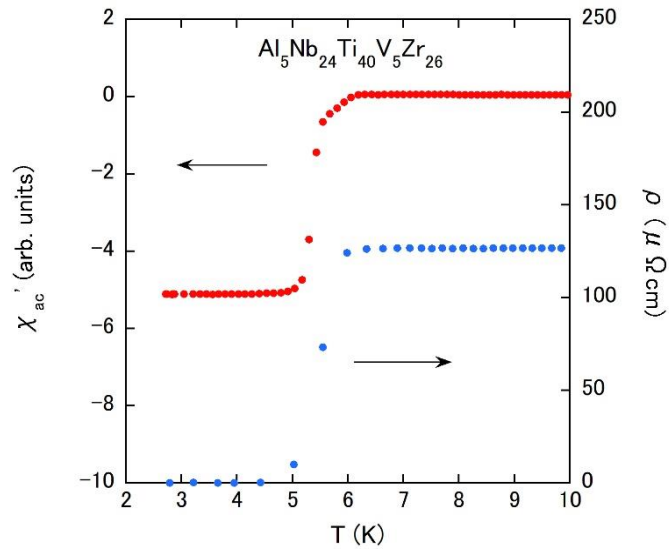

**Figure 12.** Temperature dependences of the ac magnetization and electrical resistivity of Al$_5$Nb$_{24}$Ti$_{40}$V$_5$Zr$_{26}$.

## 7. Summary

The present review updated the research status of HEA superconductors. The most investigated crystal structure is bcc, in which Hf, Zr, Ti, Ta, Nb, and V with a VEC value of 4 or 5 represent the dominant chemical components. Almost all bcc HEAs are conventional s-wave phonon-mediated Type II superconductors. Next, we surveyed the research results of hcp and hcp-related HEA superconductors. Contrary to the bcc HEA superconductors, the constituent elements are rather equally distributed on Ti, Nb, Mo, Re, Ru and Rh with different VECs. The hcp Re and Ru elements might play an important role in the stabilization of the hexagonal structure. The VEC dependence of $T_c$ for the bcc (hcp + hcp-related) HEAs is compared with those for crystalline 4$d$ metal solid solutions and amorphous 4$d$ metals. The results suggest that, in particular, bcc HEA superconductors can be regarded as an intermediate system between crystalline and amorphous superconductors. The stability of the bcc structure is tolerable for $\delta$ up to approximately 10; on the other hand, there seems to be little tolerance of $\delta$ for hcp phase stability. In the HEA superconductors of both structures, the superconducting state is robust against $\delta$. We also discussed the phase selection of bcc and hcp HEA superconductors based on e/a and VEC. The simultaneous consideration of e/a and VEC assists in reliable material design. At the present stage, the formation conditions of bcc and hcp HEA superconductors largely depend on two factors: one is the elemental makeup taking into account the VEC, and the other is the $\delta$-value, representing the atomic size mismatch between the constituents.

Recently, the concept of HEA has been extended to multi-site crystal structures, and several multi-site HEA superconductors have been reported to be CsCl-type, α-Mn, A15, NaCl-type, σ-phase and layered structures. Some HEAs show a deviation from the Matthias rule of crystalline 4$d$ metal solid solutions; however, a correlation between $T_c$ and VEC is still observed for each crystal structure type. The existence of this correlation suggests the important role of the density of states at the Fermi level determined by the crystal structure. The unexpectedly smaller $\delta$ values observed for almost all multi-site HEAs would mean that all sites contribute to the appearance of superconductivity.

Finally, we have presented the perspectives for the five topics of fcc HEA superconductors, multi-site HEA superconductors, HEA superconductors containing magnetic elements, eutectic HEA superconductors and gum metal HEA superconductors. The materials research on HEA superconductors has just begun, and we believe that there are unlimited possibilities for discovering new phenomena in this field.

**Author Contributions:** Conceptualization, J.K.; methodology, J.K.; formal analysis, J.K., S.H. and N.I.; writing—original draft preparation, J.K.; writing—review and editing, J.K.

**Acknowledgments:** J.K. is grateful for the support provided by Comprehensive Research Organization of Fukuoka Institute of Technology.

**Conflicts of Interest:** The authors declare no conflict of interest.

## References

1. Gao, M. C.; Yeh, J. -W.; Liaw, P. K.; Zhang, Y. *High-entropy alloys: fundamentals and applications*; Springer: Cham, Switzerland, **2015**.
2. Ye, Y. F.; Wang, Q.; Lu, J.; Liu, C. T.; Yang, Y. High-entropy alloy: challenges and prospects. *Mater. Today* **2016**, 19, 349-362.
3. Miracle, D. B.; Senkov, O. N. A critical review of high entropy alloys and related concepts. *Acta Mater.* **2017**, 122, 448-511.
4. Senkov, O. N.; Miracle, D. B.; Chaput, K. J.; Couzinie, J. –P. Development and exploration of refractory high entropy alloys—A review. *J. Mater. Res.* **2018**, 33, 3092-3128.
5. Yeh, J. -W.; Chen, S. -K.; Lin, S. -J.; Gan, J. -Y.; Chin, T. -S.; Shun, T. -T.; Tsau, C. -H.; Chang, S. -Y. Nanostructured high-entropy alloys with multiple principal elements: Novel alloy design concepts and outcomes. *Adv. Eng. Mater*. **2004**, 6, 299-303.
6. Otto, F.; Yang, Y.; Bei, H.; George, E. P. Relative effects of enthalpy and entropy on the phase stability of equiatomic high-entropy alloys. *Acta Mater.* **2013**, 61, 2628-2638.
7. Zhou, Y. J.; Zhang, Y.; Wang, Y. L.; Chen, G. L. Microstructure and compressive properties of multicomponent $Al_x(TiVCrMnFeCoNiCu)_{100-x}$ high-entropy alloys. *Mater. Sci. Eng. A* **2007**, 454–455, 260–265.
8. Zou, Y; Ma, H; Spolenak, R. Ultrastrong ductile and stable high-entropy alloys at small scales. *Nat. Commun.* **2015**, 6, 7748.
9. Shi, Y.; Yang, B.; Liaw, P. K. Corrosion-resistant high-entropy alloys: A review. *Metals* **2017**, 7, 43.
10. Sun, L.; Luo, Y.; Ren, X.; Gao, Z.; Du, T.; Wu, Z.; Wang, J. A multicomponent γ-type $(Gd_{1/6}Tb_{1/6}Dy_{1/6}Tm_{1/6}Yb_{1/6}Lu_{1/6})_2Si_2O_7$ disilicate with outstanding thermal stability. *Mater. Res. Lett.* **2020**, 8, 424-430.
11. Koželj, P.; Vrtnik, S.; Jelen, A.; Jazbec, S.; Jagličić, Z.; Maiti, S.; Feuerbacher, M.; Steurer, W.; Dolinšek, J. Discovery of a superconducting high-entropy alloy. *Phys. Rev. Lett*. **2014**, 113, 107001.
12. Matthias, B. T. Empirical relation between superconductivity and the number of valence electrons per atom. *Phys Rev* **1955**, 97, 74-76.
13. Collver, M. M.; Hammond, R. H. Superconductivity in "amorphous" transition-metal alloy films. *Phys Rev Lett* **1973**, 30, 92–95.
14. von Rohr, F.; Winiarski, M. J.; Tao, J.; Klimczuk, T.; Cava, R. J. Effect of electron count and chemical complexity in the Ta-Nb-Hf-Zr-Ti high-entropy alloy superconductor. *Proc. Natl. Acad. Sci.* **2016**, 113, E7144-E7150.
15. Guo, J.; Wang, H.; von Rohr, F.; Wang, Z.; Cai, S.; Zhou, Y.; Yang, K.; Li, A.; Jiang, S.; Wu, Q.; Cava, R. J.; Sun, L. Robust zero resistance in a superconducting high-entropy alloy at pressures up to 190 GPa. *Proc. Natl. Acad. Sci.* **2017**, 114, 13144-13147.
16. Sun, L.; Cava, R. J. High-entropy alloy superconductors: Status, opportunities, and challenges. *Phys. Rev. Mater.* **2019**, 3, 090301.
17. von Rohr, F. O.; Cava, R. J. Isoelectronic substitutions and aluminium alloying in the Ta-Nb-Hf-Zr-Ti high-entropy alloy superconductor. *Phys. Rev. Mater.* **2018**, 2, 034801.
18. Vrtnik, S; Koželj, P.; Meden, A.; Maiti, S.; Steurer, W; Feuerbacher, M; Dolinšek, J. Superconductivity in thermally annealed Ta-Nb-Hf-Zr-Ti high-entropy alloys. *J. Alloys Compd.* **2017**, 695, 3530-3540.
19. Zhang, X.; Winter, N.; Witteveen, C.; Moehl, T.; Xiao, Y.; Krogh, F.; Schilling, A.; von Rohr, F. O. Preparation and characterization of high-entropy alloy $(TaNb)_{1-x}(ZrHfTi)_x$ superconducting films. *Phys. Rev. Res.* **2020**, 2, 013375.
20. Kim, G.; Lee, M. –H.; Yun, J. H.; Rawat, P.; Jung, S. –G.; Choi, W.; You, T. –S.; Kim, S. J.; Rhyee, J. –S. Strongly correlated and strongly coupled s-wave superconductivity of the high entropy alloy $Ta_{1/6}Nb_{2/6}Hf_{1/6}Zr_{1/6}Ti_{1/6}$ compound. *Acta Mater.* **2020**, 186, 250-256.

21. Matsumoto, K.; Takewaki, H.; Tanaka, Y.; Miura, O.; Yamafuji, K.; Funaki, K.; Iwakuma, M.; Matsushita, T. Enhanced $J_c$ properties in superconducting NbTi composites by introducing Nb artificial pins with a layered structure. *Appl. Phys. Lett.* **1994**, 64, 115-117.
22. Yamafuji, K.; Fujiyoshi, T.; Toko, K.; Matsuno, T.; Kobayashi, T.; Kishio, K. On the magnetic field dependence of critical current density in single crystals of high-Tc superconductors. *Physica C* **1994**, 226, 133-142.
23. Nelson, W. L.; Chemey, A. T.; Hertz, M.; Choi, E.; Graf, D. E.; Latturner, S.; Albrecht-Schmitt, T. E.; Wei, K.; Baumbach, R. E. Superconductivity in a uranium containing high entropy alloy. *Sci. Rep.* **2020**, 10, 4717.
24. Wu, K. –Y.; Chen, S. –K; Wu, J. –M. Superconducting in equal molar NbTaTiZr-based high-entropy alloys. *Nat. Sci.* **2018**, 10, 110-124.
25. Marik, S.; Varghese, M.; Sajilesh, K. P.; Singh, D.; Singh, R. P. Superconductivity in equimolar Nb-Re-Hf-Zr-Ti high entropy alloy. *J. Alloys Compd.* **2018**, 769, 1059-1063.
26. Ishizu, N.; Kitagawa, J. New high-entropy alloy superconductor $Hf_{21}Nb_{25}Ti_{15}V_{15}Zr_{24}$. *Res. Phys.* **2019**, 13, 102275.
27. Pacheco, V.; Lindwall, G.; Karlsson, D.; Cedervall, J.; Fritze, S.; Ek, G.; Berastegui, P.; Sahlberg, M.; Jansson, U. Thermal stability of the HfNbTiVZr high-entropy alloy. *Inorg. Chem.* **2019**, 58, 811-820.
28. Roberts, B. W. Survey of superconductive materials and critical evaluation of selected properties. J. Phys. Chem. Ref. Data **1976**, 5, 581-821.
29. Guo, S.; Liu, C. T. Phase stability in high entropy alloys: Formation of solid-solution phase or amorphous phase. *Prog. Nat. Sci.* **2011**, 21, 433-446.
30. Lee, Y. –S.; Cava, R. J. Superconductivity in high and medium entropy alloys based on MoReRu. *Physica C* **2019**, 566, 1353520.
31. Marik, S.; Motla, K.; Varghese, M.; Sajilesh, K. P.; Singh, D.; Breard, Y.; Boullay, P.; Singh, R. P. Superconductivity in a new hexagonal high-entropy alloy. *Phys. Rev. Mater.* **2019**, 3, 060602(R)
32. Liu, B.; Wu, J.; Cui, Y.; Zhu, Q.; Xiao, G.; Wu, S.; Cao, G.; Ren, Z. Superconductivity in hexagonal Nb-Mo-Ru-Rh-Pd high-entropy alloys. *Scr. Mater.* **2020**, 182, 109-113.
33. Mizutani, U.; Sato, H.; Inukai, M.; Nishino, Y.; Zijlstra, E. S. Electrons per atom ratio determination and Hume-Rothery electron concentration rule for P-based polar compounds studied by FLAPW−Fourier calculations. *Inorg. Chem.* **2015**, 54, 930-946.
34. Mizutani, U.; Sato, H. The physics of the Hume-Rothery electron concentration rule. *Crystals* **2017**, 7, 9.
35. Stolze, K.; Tao, J.; von Rohr, F. O.; Kong, T.; Cava, R. J. Sc−Zr−Nb−Rh−Pd and Sc−Zr−Nb−Ta−Rh−Pd high-entropy alloy superconductors on a CsCl-type lattice. *Chem. Mater.* **2018**, 30, 906-914.
36. Stolze, K.; Cevallos, F. A.; Kong, T.; Cava, R. J. High-entropy alloy superconductors on an α-Mn lattice. *J. Mater. Chem. C* **2018**, 6, 10441-10449.
37. Wu, J.; Liu, B.; Cui, Y.; Zhu, Q.; Xiao, G.; Wang, H.; Wu, S.; Cao, G.; Ren, Z. Polymorphism and superconductivity in the V-Nb-Mo-Al-Ga high-entropy alloys. *Sci. China Mater.* **2020**, 63, 823-831.
38. Izumi, F.; Momma, K. Three-dimensional visualization in powder diffraction. *Solid State Phenom.* **2007**, 130, 15-20.
39. Tsubota, M.; Kitagawa, J. A necessary criterion for obtaining accurate lattice parameters by Rietveld method. *Sci. Rep.* **2017**, 7, 15381.
40. Mizuguchi, Y. Superconductivity in high-entropy-alloy telluride $AgInSnPbBiTe_5$. *J. Phys. Soc. Jpn.* **2019**, 88, 124708.
41. Kasem, M. R.; Hoshi, K.; Jha, R.; Katsuno, M.; Yamashita, A.; Goto, Y.; Matsuda, T. D.; Aoki, Y.; Mizuguchi, Y. Superconducting properties of high-entropy-alloy tellurides M-Te (M: Ag, In, Cd, Sn, Sb, Pb, Bi) with a NaCl-type structure. *Appl. Phys. Express* **2020**, 13, 033001.
42. Yamashita, A.; Jha, R.; Goto, Y.; Matsuda, T. D.; Aoki, Y.; Mizuguchi, Y. An Efficient way of increasing the total entropy of mixing in high-entropy-alloy compounds: a case of NaCl-type (Ag,In,Pb,Bi)$Te_{1-x}Se_x$ ($x = 0.0$, 0.25, 0.5) superconductors. *Dalton Trans.* **2020**, 49, 9118-9122.
43. Liu,B.; Wu, J.; Cui, Y.; Zhu, Q.; Xiao, G.; Wang, H.; Wu, S.; Cao, G.; Ren, Z. Formation and superconductivity of single-phase high-entropy alloys with a tetragonal structure. *ACS Appl. Electron. Mater.* **2020**, 2, 1130-1137.
44. Sogabe, R.; Goto, Y.; Mizuguchi, Y. Superconductivity in $REO_{0.5}F_{0.5}BiS_2$ with high entropy-alloy-type blocking layers. *Appl. Phys. Express* **2018**, 11, 053102.

45. Sogabe, R.; Goto, Y.; Abe, T., Moriyoshi, C.; Kuroiwa, Y.; Miura, A.; Tadanaga, K.; Mizuguchi, Y. Improvement of superconducting properties by high mixing entropy at blocking layers in BiS$_2$-based superconductor REO$_{0.5}$F$_{0.5}$BiS$_2$. *Solid State Commun.* **2019**, 295, 43-49.
46. Shukunami, Y.; Yamashita, A.; Goto, Y.; Mizuguchi, Y. Synthesis of RE123 high-$T_c$ superconductors with a high-entropy-alloy-type RE site. *Physica C*, **2020**, 572, 1353623.
47. Islam, N.; Huang, W.; Zhuang, H. L. Machine learning for phase selection in multi-principal element alloys. *Comput. Mater. Sci.* **2018**, 150, 230-235.
48. Li, Y.; Guo, W. Machine-learning model for predicting phase formations of high-entropy alloys. *Phys. Rev. Mater.* **2019**, 3, 095005.
49. Ishizu, N.; Kitagawa, J. Trial of a search for a face-centered-cubic high-entropy alloy superconductor. arXiv:2007.00788 **2020**.
50. Cort, B.; Giorgi, A. L.; Stewart, G. R. Low temperature specific heats of H(NbIrO) and R(NbPtO). *J. Low. Temp. Phys.* **1982**, 47, 179-185.
51. Waterstrat, R. M.; Kuentzler, R.; Muller, J. Structural instabilities and superconductivity in quasi-binary Mn$_5$Si$_3$-type compounds. *J. Less Common Met.* **1990**, 167, 169-178.
52. Lv, B.; Zhu, X. Y.; Lorenz, B.; Wei, F. Y.; Xue, Y. Y.; Yin, Z. P.; Kotliar, G.; Chu, C. W. Superconductivity in the Mn$_5$Si$_3$-type Zr$_5$Sb$_3$ system. *Phys. Rev. B* **2013**, 88, 134520.
53. Zhang, Y.; Wang, B.; Xiao, Z.; Lu, Y.; Kamiya, T.; Uwatoko, Y.; Kageyama, H.; Hosono, H. Electride and superconductivity behaviors in Mn$_5$Si$_3$-type intermetallics. *npj Quantum Materials* **2017**, 2, 45.
54. Li, S.; Liu, X.; Anand, V.; Lv, B. Superconductivity from site-selective Ru doping studies in Zr$_5$Ge$_3$ compound. *New J. Phys.* **2018**, 20, 013009.
55. Hamamoto, S.; Kitagawa, J. Superconductivity in oxygen-added Zr$_5$Pt$_3$. *Mater. Res. Express* **2018**, 5, 106001.
56. Xu, Y.; Jöhr, S.; Das, L.; Kitagawa, J.; Medarde, M.; Shiroka, T.; Chang, J.; Shang. T. Crossover from multiple- to single-gap superconductivity in Nb$_5$Ir$_{3-x}$Pt$_x$O alloys. *Phys. Rev. B* **2020**, 101, 134513.
57. Corbett, J. D.; Garcia, E.; Guloy, A. M.; Hurng, W. -M. Kwon, Y. -U. Leon-Escamilla. E. A. Widespread interstitial chemistry of Mn$_5$Si$_3$-type and related phases. Hidden impurities and opportunities. *Chem. Mater.* **1998**, 10, 2824-2836.
58. Kitagawa, J.; Hamamoto, S. Superconductivity in Nb$_5$Ir$_{3-x}$Pt$_x$O. *JPS Conf. Proc.* **2020**, 30, 011055.
59. Ishikawa, M. Can magnetism and superconductivity coexist? *Contemp. Phys.* **1982**, 23, 443-468.
60. Rogacki, K.; Batlogg, B.; Karpinski, J.; Zhigadlo, N. D.; Schuck, G.; Kazakov, S. M.; Wägli, P.; Puźniak, R.; Wiśniewski, A.; Carbone, F.; Brinkman, A.; van der Marel, D. Strong magnetic pair breaking in Mn-substituted MgB$_2$ single crystals. *Phys. Rev. B* **2006**, 73, 174520.
61. Lu, Y.; Dong, Y.; Guo, S.; Jiang, L.; Kang, H.; Wang, T.; Wen, B.; Wang, Z.; Jie, J.; Cao, Z.; Ruan, H.; Li, T. A promising new class of high-temperature alloys: Eutectic high-entropy alloys. *Sci. Rep.* **2015**, 4, 6200.
62. Maeno, Y.; Ando, T.; Mori, Y.; Ohmichi, E.; Ikeda, S.; Nishizaki, S.; Nakatsuji, S. Enhancement of Superconductivity of Sr$_2$RuO$_4$ to 3 K by Embedded Metallic Microdomains. *Phys. Rev. Lett.* **1998**, 81, 3765-3768.
63. Matthias, B. T.; Stewart, G. R.; Giorgi, A. L.; Smith, J. L.; Fisk, Z.; Barz, H. Enhancement of superconductivity through lattice softening. *Science* **1980**, 208, 401-402.
64. Cline, H. E.; Rose, R. M.; Wulff, J. Niobium-thorium eutectic alloy as a high-field, high-current superconductor. *J. Appl. Phys.* **1963**, 34, 1771-1774.
65. Xu, Q.; Chen, D.; Tan, C.; Bi, X.; Wang, Q.; Cui, H.; Zhang, S.; Chen, R. NbMoTiVSi$_x$ refractory high entropy alloys strengthened by forming BCC phase and silicide eutectic structure. *J. Mater. Sci. Technol.* **2021**, 60, 1-7.
66. Hafstrom, J. W. Metallurgy, fabrication, and superconducting properties of multifilamentary Nb$_3$Al composites. *IEEE Trans. Magn.* **1977**, MAG-13, 480-482.
67. Saito, T.; Furuta, T.; Hwang, J. -H.; Kuramoto, S.; Nishino, K.; Suzuki, N.; Chen, R.; Yamada, A.; Ito, K.; Seno, Y.; Nonaka, T.; Ikehata, H.; Nagasako, N.; Iwamoto, C.; Ikuhara, Y.; Sakuma, T. Multifunctional alloys obtained via a dislocation free plastic deformation mechanism. *Science* **2003**, 300, 464–467.
68. Zherebtsov, S.; Yurchenko, N.; Panina, E.; Tikhonovsky, M.; Stepanov, N. Gum-like mechanical behavior of a partially ordered Al$_5$Nb$_{24}$Ti$_{40}$V$_5$Zr$_{26}$ high entropy alloy. *Intermetallics* **2020**, 116, 106652.